\documentclass[prd,twocolumn,reprint,superscriptaddress,showpac]{revtex4-2}

\usepackage{feynmp,amsmath,amssymb,graphicx,subfigure,hyperref,bm,mathrsfs,slashed,color,CJK}

\begin{document}
	\begin{CJK}{UTF8}{song}
		
		\title{Effects of spatial dimensionality and band tilting on the longitudinal optical conductivities in Dirac bands}

		\author{Jian-Tong Hou}
		\thanks{These authors have contributed equally to this work.}
		\affiliation{Department of Physics, Institute of Solid State Physics and Center for Computational Sciences,
			Sichuan Normal University, Chengdu, Sichuan 610066, China}
		\affiliation{College of Physics, Sichuan University, Chengdu, Sichuan 610064, China}
		
		\author{Chang-Xu Yan}
		\thanks{These authors have contributed equally to this work.}
		\affiliation{Department of Physics, Institute of Solid State Physics and Center for Computational Sciences,
			Sichuan Normal University, Chengdu, Sichuan 610066, China}
		
		\author{Chao-Yang Tan}
		\affiliation{Department of Physics, Institute of Solid State Physics and Center for Computational Sciences,
			Sichuan Normal University, Chengdu, Sichuan 610066, China}
		
		\author{Zhi-Qiang Li}
		\affiliation{College of Physics, Sichuan University, Chengdu, Sichuan 610064, China}

		\author{Peng Wang}
		\affiliation{College of Physics, Sichuan University, Chengdu, Sichuan 610064, China}
		\affiliation{Center for Theoretical Physics, Sichuan University, Chengdu, 610064, China}
		
		\author{Hong Guo}
		\thanks{Corresponding author:hong.guo@mcgill.ca}
		\affiliation{Department of Physics, McGill University, Montreal, Quebec H3A 2T8, Canada}
		\affiliation{Department of Physics, Institute of Solid State Physics and Center for Computational Sciences,
			Sichuan Normal University, Chengdu, Sichuan 610066, China}

		\author{Hao-Ran Chang}
		\thanks{Corresponding author:hrchang@mail.ustc.edu.cn}
		\affiliation{Department of Physics, Institute of Solid State Physics and Center for Computational Sciences,
			Sichuan Normal University, Chengdu, Sichuan 610066, China}
		\affiliation{College of Physics, Sichuan University, Chengdu, Sichuan 610064, China}
		\affiliation{Department of Physics, McGill University, Montreal, Quebec H3A 2T8, Canada}
		
		\date{\today}
		
		\begin{abstract}
			We report a unified theory based on linear response, for analyzing the longitudinal optical conductivity (LOC) of materials with tilted Dirac cones. Depending on the tilt parameter $t$, the Dirac electrons have four phases: untilted, type-I, type-II, and type-III; the Dirac dispersion can be isotropic or anisotropic; the spatial dimension of the material can be one-, two-, or three-dimensions (1D, 2D and 3D).
			The interband LOCs and intraband LOCs in $d$ dimension (with $d\ge2$) are found to scale as $\sigma_{0}\omega^{d-2}$ and $\sigma_{0}\mu^{d-1}\delta(\omega)$, respectively, where $\omega$ is the frequency and $\mu$ the chemical potential. The interband LOC vanishes in 1D due to lack of extra spatial dimension. In contrast, the interband LOCs in 2D and 3D are nonvanishing and share many similar properties. A universal and robust fixed point of interband LOCs appears at $\omega=2\mu$ no matter $d=2$ or $d=3$, which can be intuitively understood by the geometric structures of Fermi surface and energy resonance contour. The intraband LOCs and the carrier density for 2D and 3D tilted Dirac bands are both closely related to the geometric structure of Fermi surface and the cutoff of integration. The angular dependence of LOCs is found to characterize both spatial dimensionality and band tilting and the constant asymptotic background values of LOC reflect features of Dirac bands.
			The LOCs in the anisotropic tilted Dirac cone can be connected to its isotropic counterpart by a ratio that consists of Fermi velocities for both 2D and 3D. Most of the findings are universal for tilted Dirac materials and hence valid for a great many Dirac materials in the spatial dimensions of physical interest.
		\end{abstract}
		\maketitle
	\end{CJK}

	\section{Introduction\label{Sec:intro}}
	
	Dirac fermions in condensed matter materials have attracted great and continued attention since the exfoliation of single layer graphene \cite{Science2004}.
	The low energy electronic structure of Dirac fermions is characterized by the one-, two-, and three-dimensional (1D, 2D, and 3D) Dirac cones in the momentum space \cite{RMP2009,RMP2018,SatoPRB2018,Yangnjp2020,Zhunjp2020,WangACSNano2021,Zhangnjp2022,YueNanoLett2022} and may be categorized into four distinct ``phases" via a tilt parameter
	$t$, namely the untilted phase ($t=0$), type-I phase ($0<t<1$), type-II phase ($t>1$), and type-III phase ($t=1$) \cite{JPSJ2006,Zhou8Pmmn2014PRL,
		Science8Pmmn2015,Nature2015,Volovik2017,Volovik2018,Liu2021}. Here, the tilt indicates how the Dirac cone is oriented in the momentum space and pristine graphene belongs
	to the untilted category. The behavior of experimentally measurable physical quantities in these phases of Dirac fermions are important to understand, and may be
	used as detection signals for the intrinsic electronic structure of the Dirac material. To this end, the longitudinal optical conductivity (LOC) of Dirac electrons, which are remarkably different from that of electron gases\cite{AndoRMP1982,SharmaIJMPB2002,MaslovPRB2015, MawrieJPCM2016,VermaPRB2020}, is particularly interesting.

	Earlier works on the LOC have focused mostly on the untilted Dirac materials in 2D and 3D, including graphene \cite{PRLCarbotte2006,PRBGusynin2007,PRLMikhailov2007,PRLMarel2008,PRLMak2008,
		PRBStauber2008}, silicene \cite{PRBStille2012}, and Dirac/Weyl semimetals \cite{PRBAshby2014}. Thereafter, theoretical works on LOC in the 2D and 3D Dirac materials were extended to type-I \cite{JPSJNishine2010,PRBVerma2017,PRBHerrera2019,PRBGoerbig2019,
		PRBTan2021,PRBJDOS2021,PRBMojarro2022}, type-II \cite{PRBWild2022,PRBTan2022,PRBCarbotte2016,PRBCarbotte2017}, and type-III phases \cite{PRBWild2022,PRBTan2022}. Refs. \cite{PRBCarbotte2016,PRBCarbotte2017} reported
	interesting effects of Dirac cone tilting on the optical response of type-I and type-II 3D Weyl semimetals, although type-III phase was unexplored. More recently, References \cite{PRBWild2022,PRBTan2022} suggested that the LOCs in 2D tilted Dirac materials differ dramatically among untilted, type-I, type-II, and type-III phases.
	
	The rather complicated and multitude properties of LOC, in different phases of tilted Dirac bands and in different spatial dimensions (1D, 2D, and 3D), call for a unified comprehensive theoretical analysis. In condensed phase materials, spatial dimensionality and energy dispersion of carriers play central roles in determining physical properties, as exemplified by well-known situations such as phase transitions, localization, Klein tunneling, topologically protected helical states, etc. It is the purpose of this work to provide such a unified analysis of LOC against spatial dimension and band dispersion. Our comparable study not only reports qualitative differences/similarities of different spatial dimensionality and band tilting, but also reveals several robust properties independent of spatial dimensionality.
	
	For materials with tilted Dirac bands, our analysis discovered two general scale relations for the interband and intraband LOCs. The interband LOC vanishes in 1D for lack of extra spatial dimension. By contrast, the nonvanishing interband LOCs in 2D and 3D exhibit many similar properties. A universal and robust fixed point of interband LOCs is found at $\omega=2\mu$, no matter in 2D or 3D, which is intuitively analyzed from the geometric structure of Fermi surface and energy resonance contour. The intraband LOCs and the carrier density for 2D and 3D tilted Dirac bands are both closely related to the geometric structure of Fermi surface and the cutoff of integration. The angular dependence of LOC is found to provide a means for characterizing both spatial dimensionality and band tilting and the constant asymptotic background values of LOC reflect the essential features of tilted Dirac bands. The LOCs in the anisotropic tilted Dirac cone can be connected to its isotropic counterpart by a ratio that consists of Fermi velocities of the individual material. The conclusions of this analysis are valid for a great many Dirac materials in all spatial dimensions of physical interests
		\cite
		{ RMP2009,RMP2018,SatoPRB2018,Yangnjp2020,Zhunjp2020,WangACSNano2021,Zhangnjp2022,YueNanoLett2022,JPSJ2006,Zhou8Pmmn2014PRL,Science8Pmmn2015,Nature2015,Volovik2017,Volovik2018,Liu2021,PRLCarbotte2006,PRBGusynin2007,PRLMikhailov2007,PRLMarel2008,PRLMak2008,PRBStauber2008,PRBStille2012,PRBAshby2014,JPSJNishine2010,PRBVerma2017,PRBHerrera2019,PRBGoerbig2019,PRBTan2021,PRBJDOS2021,PRBMojarro2022,PRBWild2022,PRBTan2022}.
	
	The rest of this paper is organized as follows. In Sec.{\ref{Sec:Theoretical formalism}}, we briefly describe the model Hamiltonian for \emph{isotropic} tilted Dirac fermions and the theoretical formalism to calculate the LOCs. The analytical expressions of the interband and intraband (Drude) LOCs are presented in Sec.{\ref{Sec:Interband Part}} and Sec.{\ref{Sec:Intraband Part}}, respectively. The angular dependence is shown in Sec. {\ref{Sec:Angular dependence}}. 
	We discuss the effect of anisotropy based on the variants of an \emph{isotropic} model Hamiltonian in Sec.{\ref{Sec:Effect of anisotropy}}. Our main conclusions are summarized in Sec.{\ref{Sec:Summary}}. Finally, we provide three appendices to present detailed calculations and analysis.
	
	\section{Model and Theoretical formalism \label{Sec:Theoretical formalism}}
	
	We begin with the low energy Hamiltonian of the \emph{isotropic} tilted Dirac fermions
	\begin{align}
		\mathcal{H}_{\kappa}(\boldsymbol{k})=\kappa \hbar v_tk_{1}\tau_{0}+\hbar
		v_{F}\boldsymbol{k}\cdot\boldsymbol{\tau},\label{Eq1}
	\end{align}
	where $\kappa=\pm$ denotes the valley of Dirac point, and $\boldsymbol{k}=(k_1,\cdots,k_d)$ represents the wave vector in $d$ dimensions. Hereafter we set $\hbar=v_F=1$ for simplicity, focus on the spatial dimension of physical interest with $d=1,2,3$
	and relabel Cartesian indices $j=x,y,z$ as $j=1,2,3$. We introduce the tilt parameter $t=v_t/v_F$ to account for the band tilting along the first direction of spatial dimension $k_1$. As mentioned in the Introduction, in general Dirac materials can be classified into four distinct phases: untilted phase ($t=0$), type-I phase ($0<t<1$), type-II phase ($t>1$), and type-III phase ($t=1$). In addition, the $2\times2$ matrices $\tau_0$ and $\tau_j$ stand for the identical matrix and the $j$-th Pauli matrix operating in the pseudospin space, respectively. The eigenvalues of the Hamiltonian (\ref{Eq1}) can be straightforwardly obtained,
	\begin{align}
		\varepsilon^{\lambda}_{\kappa}(\boldsymbol{k})=\kappa t  k_{1}+\lambda|\boldsymbol{k}|,\label{Eq2}
	\end{align}
	where $\lambda=\pm$ are the conduction band and valence band, respectively. Within linear response theory, the LOC at a finite photon frequency $\omega$ is given by
	\begin{align}
		\sigma_{jj}(\omega,d)=g_s\sum_{\kappa=\pm}\sigma_{jj}^{\kappa}(\omega,d),
		\label{Eq3}
	\end{align}
	where $g_s=2$ gives spin degeneracy.
	
	\begin{widetext}
		The LOC at a given valley $\kappa$ are generally written as \cite{Vignale2005,Mahan2007,PRBTan2021,PRBTan2022}
			\begin{align}\label{SigmaDef}
				\sigma_{jj}^{\kappa}(\omega,d)=\frac{\text{i}}{\omega}\lim_{\boldsymbol{q}\to\boldsymbol{0}}
				 \int_{-\infty}^{+\infty}\frac{\mathrm{d}^{d}\boldsymbol{k}}{(2\pi)^{d}}\sum_{\lambda,\lambda^{\prime}=\pm}
				\mathcal{F}_{\lambda,\lambda^{\prime}}^{\kappa;jj}(\boldsymbol{k},\boldsymbol{k}+\boldsymbol{q})
				 \frac{f[\varepsilon^{\lambda}_{\kappa}(\boldsymbol{k})]-f[\varepsilon_{\kappa}^{\lambda^{\prime}}(\boldsymbol{k}+\boldsymbol{q})]}{\omega+\varepsilon^{\lambda}_{\kappa}(\boldsymbol{k})-\varepsilon_{\kappa}^{\lambda^{\prime}}(\boldsymbol{k}+\boldsymbol{q})+\text{i}\eta},
			\end{align}
			where
			\begin{align}\label{Ffactor}
				\mathcal{F}_{\lambda,\lambda^{\prime}}^{\kappa;jj}(\boldsymbol{k},\boldsymbol{k}+\boldsymbol{q})&=
				\frac{e^{2}}{2}\left\{ t^{2}\delta_{j1}\delta_{j1}\left[1+\lambda\lambda'\frac{\boldsymbol{k}\cdot(\boldsymbol{k}+\boldsymbol{q})}{|\boldsymbol{k}||\boldsymbol{k}+\boldsymbol{q}|}\right]
				+2\kappa t
				 \delta_{j1}\left[\lambda\frac{k_{j}}{|\boldsymbol{k}|}+\lambda'\frac{k_{j}+q_{j}}{|\boldsymbol{k}+\boldsymbol{q}|}\right]\right\}
				\nonumber\\&
				 +\frac{e^{2}}{2}\left\{\delta_{jj}\left[1-\lambda\lambda'\frac{\boldsymbol{k}\cdot(\boldsymbol{k}+\boldsymbol{q})}{|\boldsymbol{k}||\boldsymbol{k}+\boldsymbol{q}|}\right]+2\lambda\lambda'\frac{k_{j}(k_{j}+q_{j})}{|\boldsymbol{k}||\boldsymbol{k}+\boldsymbol{q}|}\right\}
			\end{align}
			and $\eta$ denotes a positive infinitesimal. Accordingly, the real part of LOC at a given valley $\kappa$ can be divided into an interband part and an intraband/Drude part as
		\begin{align}\label{Eq.6}
			\mathrm{Re}\sigma_{jj}^{\kappa}(\omega,d)=\mathrm{Re}\sigma_{jj}^{\kappa(\mathrm{IB})}(\omega,d)
			+\mathrm{Re}\sigma_{jj}^{\kappa(\mathrm{D})}(\omega,d),
		\end{align}
		where
		\begin{align}\label{Eq.7}
			\mathrm{Re}\sigma_{jj}^{\kappa(\mathrm{IB})}(\omega,d)
			&=\pi\int_{-\infty}^{+\infty}\frac{\mathrm{d}^{d}\boldsymbol{k}}{(2\pi)^{d}}
			\mathcal{F}_{-,+}^{\kappa;jj}(\boldsymbol{k})
			\frac{f\left[\varepsilon_{\kappa}^{-}(\boldsymbol{k})\right]
				-f\left[\varepsilon_{\kappa}^{+}(\boldsymbol{k})\right]}{\omega}
			\delta\left(\omega-2|\boldsymbol{k}|\right)
			\notag\\&=
			4\pi\sigma_{0}\int_{-\infty}^{+\infty}\frac{\mathrm{d}^{d}\boldsymbol{k}}{(2\pi)^{d}}
			\left[\delta_{jj}-\frac{k_jk_j}{|\boldsymbol{k}|^2}\right]
			\frac{f\left[\varepsilon_{\kappa}^{-}(\boldsymbol{k})\right]
				-f\left[\varepsilon_{\kappa}^{+}(\boldsymbol{k})\right]}{\omega}
			\delta\left(\omega-2|\boldsymbol{k}|\right),
		\end{align}
		and
		\begin{align}\label{Eq.8}
			\mathrm{Re}\sigma_{jj}^{\kappa(\mathrm{D})}(\omega,d)
			&=\pi\sum_{\lambda=\pm}\int_{-\infty}^{+\infty}\frac{\mathrm{d}^{d}\boldsymbol{k}}{(2\pi)^{d}}
			\mathcal{F}_{\lambda,\lambda}^{\kappa;jj}(\boldsymbol{k})
			 \left\{-\frac{\mathrm{d}f\left[\varepsilon^{\lambda}_{\kappa}(\boldsymbol{k})\right]}{\mathrm{d}\varepsilon^{\lambda}_{\kappa}(\boldsymbol{k})}\right\}
			\delta(\omega)
			\notag\\&=
			4\pi\sigma_{0}\sum_{\lambda=\pm}\int_{-\infty}^{+\infty}\frac{\mathrm{d}^{d}\boldsymbol{k}}{(2\pi)^{d}}
			\frac{\left(\kappa t |\boldsymbol{k}|\delta_{j1}+\lambda k_j\right)
				\left(\kappa t |\boldsymbol{k}|\delta_{j1}+\lambda k_j\right)}{|\boldsymbol{k}|^2}
			 \left\{-\frac{\mathrm{d}f\left[\varepsilon^{\lambda}_{\kappa}(\boldsymbol{k})\right]}{\mathrm{d}\varepsilon^{\lambda}_{\kappa}(\boldsymbol{k})}\right\}
			\delta(\omega),
		\end{align}
	with
			\begin{align}\label{Eq.9}
				\mathcal{F}_{\lambda,\lambda^{\prime}}^{\kappa;jj}(\boldsymbol{k})\equiv\lim_{\boldsymbol{q}\to 0}	 \mathcal{F}_{\lambda,\lambda^{\prime}}^{\kappa;jj}(\boldsymbol{k},\boldsymbol{k}+\boldsymbol{q})
				 =e^{2}(1-\delta_{\lambda\lambda^{\prime}})\left[\delta_{jj}-\frac{k_{j}k_{j}}{|\boldsymbol{k}|^{2}}\right]+e^{2}\delta_{\lambda\lambda^{\prime}}\frac{\left(\kappa t|\boldsymbol{k}|\delta_{j1}+\lambda k_{j}\right)\left(\kappa t|\boldsymbol{k}|\delta_{j1}+\lambda k_{j}\right)}{|\boldsymbol{k}|^{2}}.
			\end{align}
			In Eq.(\ref{Eq.9}), the first term contributes to the interband transition ($\lambda^{\prime}\neq\lambda$), while the second term accounts for the intraband transition ($\lambda^{\prime}=\lambda$). Especially in 1D, the interband LOC always vanishes due to $\left[\delta_{jj}-\frac{k_jk_j}{|\boldsymbol{k}|^2}\right]\equiv0$ therein, indicating
			a significantly qualitative difference between 1D and higher dimensions.
			It is noted that the detailed derivation from Eqs.(\ref{SigmaDef}) and (\ref{Ffactor}) to Eqs.(\ref{Eq.7}) - (\ref{Eq.9}) can be found in Appendix {\ref{apdx.01}}.
	\end{widetext}
	
	In these definitions, $\sigma_0=\frac{e^2}{4\hbar}$ (we restore $\hbar$ for explicitness), $\delta(x)$ is the Dirac delta function
	and the Fermi distribution function $f(x)=\left\{1+\exp[(x-\mu)/k_BT]\right\}^{-1}$
	in which $\mu$ measures the chemical potential with respect to the Dirac point, $k_B$ is the Boltzmann constant, and $T$ denotes the temperature. We assume zero temperature $T\to0$ in order to obtain an analytical expression such that the Fermi distribution function $f(x)$ and its derivative $-f'(x)$ turn out to be the Heaviside step function $\Theta(\mu-x)$ and the delta function $\delta(x-\mu)$, respectively. A similar procedure yields the particle-hole symmetry \cite{PRBTan2022}, indicating that the LOC depends on $|\mu|$, rather than $\mu$; hence we restrict to the case $\mu\ge 0$ hereafter.

	After taking the contribution of both valleys into account, the real part of LOCs can be recast as
	\begin{align}
		\mathrm{Re}\sigma_{jj}(\omega,d)
		&=\mathrm{Re}\sigma_{jj}^{(\mathrm{IB})}(\omega,d)+\mathrm{Re}\sigma_{jj}^{(\mathrm{D})}(\omega,d),
	\end{align}
	where
	$\mathrm{Re}\sigma_{jj}^{(\mathrm{IB})}(\omega,d)$ and $\mathrm{Re}\sigma_{jj}^{(\mathrm{D})}(\omega,d)$ denote the intraband LOCs and interband LOCs, respectively.

	\section{Interband conductivity \label{Sec:Interband Part}}
	The real part of interband LOCs can be generally written to be
	\begin{align}
		\mathrm{Re}\sigma_{jj}^{(\mathrm{IB})}(\omega,d)
		=S_{jj}^{(\mathrm{IB})}(\omega,d)~\Gamma_{jj}^{(\mathrm{IB})}(\omega,d;\mu, t ),\label{IBjj}
	\end{align}
	where the dimensionless functions $\Gamma_{jj}^{(\mathrm{IB})}(\omega,d;\mu, t )$ characterize the interband
	LOCs for tilted Dirac cone and the dimension-dependent magnitudes $S_{jj}^{(\mathrm{IB})}(\omega,d)$ are defined
	as the undoped ($\mu=0$) interband LOCs for untilted Dirac cone ($ t =0$), namely,
	\begin{align}\label{Eq.11}
		&S_{jj}^{(\mathrm{IB})}(\omega,d)
		=\left[\mathrm{Re}\sigma_{jj}^{(\mathrm{IB})}(\omega,d)\right]_{\mu= t =0},
	\end{align}
	which is the dimension-dependent magnitude, reflecting the scaling of dimensional dependence with respect to
	frequency $\omega$ in the interband LOCs.
	
	The explicit definition of $S_{jj}^{(\mathrm{IB})}(\omega,d)$ is given by
	\begin{align}
		&\hspace{-0.3cm}S_{jj}^{(\mathrm{IB})}(\omega,d)
		\notag\\&\hspace{-0.3cm}=
		4\pi g_sg_v\sigma_0\int_{-\infty}^{+\infty}\frac{\mathrm{d}^{d}\boldsymbol{k}}{(2\pi)^{d}}
		\left[\delta_{jj}-\frac{k_jk_j}{|\boldsymbol{k}|^2}\right]
		\frac{\delta\left(\omega-2|\boldsymbol{k}|\right)}{\omega},
	\end{align}
	which can be further written as
	\begin{align}
		S_{jj}^{(\mathrm{IB})}(\omega,d)
		&=
		\begin{cases}
			S_{\parallel}^{(\mathrm{IB})}(\omega,d), & j=1,\\\\
			S_{\perp}^{(\mathrm{IB})}(\omega,d), & j\neq 1,
		\end{cases}
	\end{align}
	with $g_v=2$ denoting the valley degeneracy.
	
	We note a simple fact that the spatial components in 2D (3D) can be $j=1,2$ ($j=1,2,3$), while the spatial component in 1D has no other choice than $j=1$. It can be found that $S_{\chi}^{(\mathrm{IB})}(\omega,d)$ in 1D differs qualitatively from that in 2D and 3D. With respect to the tilting direction (the first direction $j=1$), the only permitted $S_{\chi}^{(\mathrm{IB})}(\omega,d)$ in 1D is $S_{\parallel}^{(\mathrm{IB})}(\omega,d=1)$ along the parallel direction due to the lack of extra spatial dimensionality; in contrast $S_{\parallel}^{(\mathrm{IB})}(\omega,d)$ and $S_{\perp}^{(\mathrm{IB})}(\omega,d)$ are well-defined in 2D and 3D. A
	more striking result is that the interband LOC for the 1D tilted Dirac band always vanishes, as a direct consequence of $S_{\parallel}^{(\mathrm{IB})}(\omega,d=1)\equiv0$. This originates from the fact that the relation $\mathcal{F}_{\lambda,\lambda^{\prime}}^{\kappa;jj}(\boldsymbol{k})=0$ always holds for 1D when $\textcolor[rgb]{1.00,0.00,0.00}{\lambda^{\prime}\neq \lambda}$ (interband transition), showing once again the significant qualitative difference between 1D and higher dimensions.
	
	In 2D and 3D, both $S_{\parallel}^{(\mathrm{IB})}(\omega,d)$ and $S_{\perp}^{(\mathrm{IB})}(\omega,d)$ satisfy
	\begin{align}
		S_{\chi}^{(\mathrm{IB})}(\omega,d)\equiv S^{(\mathrm{IB})}(\omega,d),
	\end{align}
	with $\chi=\parallel,\perp$. A general relation $S^{(\mathrm{IB})}(\omega,d)\propto\sigma_{0}
	\omega^{d-2}$ can be obtained from dimensional analysis for $d\geq2$. This result yields the same power dependence of interband LOCs with respect to $\omega$ in Refs. \cite{PRBCarbotte2016,PRBTan2022}. Specifically, the explicit expressions of $S^{(\mathrm{IB})}(\omega,d)$ for $d=1,2,3$ are tabulated in Table \ref{Tab1}.
	\begin{table}[htbp]
		\begin{tabular*}
			{\columnwidth}{@{\extracolsep{\fill}} c | c c c}
			\hline
			\hline
			$ d $  &$1$ &$2$ &$3$
			\tabularnewline
			\hline
			$S^{(\mathrm{IB})}(\omega,d)$  &0 &$\sigma_0$ & $\frac{1}{v_{F}}\frac{2}{3\pi }\sigma_{0}\omega$
			\tabularnewline
			\hline
			\hline
		\end{tabular*}
		\caption{Explicit expressions of $S^{(\mathrm{IB})}(\omega,d)$ for $d=1,2,3$. Here we restore $v_F$ for explicitness.}
		\label{Tab1}
	\end{table}
	
	The dimensionless functions $\Gamma_{jj}^{(\mathrm{IB})}(\omega,d;\mu, t )$ can be classified to be
	\begin{align}
		\Gamma_{jj}^{(\mathrm{IB})}(\omega,d;\mu, t )
		&=
		\begin{cases}
			\Gamma_{\parallel}^{(\mathrm{IB})}(\omega,d;\mu, t ), & j=1,\\\\
			\Gamma_{\perp}^{(\mathrm{IB})}(\omega,d;\mu, t ), & j\neq 1.
		\end{cases}
	\end{align}
	By utilizing these definitions, we recast $\mathrm{Re}\sigma_{jj}^{(\mathrm{IB})}(\omega,d)$ as
	\begin{align}
		\mathrm{Re}\sigma_{jj}^{(\mathrm{IB})}(\omega,d)=
		\begin{cases}
			\mathrm{Re}\sigma_{\parallel}^{(\mathrm{IB})}(\omega,d), & j=1,\\\\
			\mathrm{Re}\sigma_{\perp}^{(\mathrm{IB})}(\omega,d), & j\neq 1,
		\end{cases}
	\end{align}
	where
	\begin{align}
		\mathrm{Re}\sigma_{\chi}^{(\mathrm{IB})}(\omega,d)
		=S_{\chi}^{(\mathrm{IB})}(\omega,d)\;\Gamma_{\chi}^{(\mathrm{IB})}(\omega,d;\mu, t ),\label{Eq7}
	\end{align}
	with $\chi=\parallel,\perp$. Since $S^{(\mathrm{IB})}(\omega,d=1)=0$, it is not necessary to calculate $\Gamma_{jj}^{(\mathrm{IB})}(\omega,d;\mu,t)$ for the 1D tilted Dirac cone. In the following subsections, we present analytical expressions of $\Gamma_{\chi}^{(\mathrm{IB})}(\omega,d;\mu,t)$ in 2D and 3D for
	all tilted phases ($t=0$, $0<t<1$, $t=1$, and $t>1$) and both components ($\chi=\parallel$ and $\chi=\perp$).
	
	\subsection{Unified expressions of $\Gamma_{\chi}^{(\mathrm{IB})}(\omega,d;\mu, t )$}
	
	To simplify results, we introduce two compact notations
		\begin{align}
			\omega_{\pm}(t)&=\frac{2\mu}{t\mp 1}, \\
			\zeta_{\pm}(\omega)&=\frac{\omega\pm 2\mu}{\omega}\frac{\tilde{\Theta}(t)}{t},
		\end{align}
		where $\tilde{\Theta}(x)=0$ for $x<0$ and $\tilde{\Theta}(x)=1$ for $x\geq0$, and four auxiliary functions
		\begin{align}\label{Eq.14}
			\mathcal{G}_{\parallel}^{(2)}(x)&=\frac{1}{\pi}\arcsin x+\frac{x}{\pi}\sqrt{1-x^{2}},\\
			\mathcal{G}_{\perp}^{(2)}(x)&=\frac{1}{\pi}\arcsin x-\frac{x}{\pi}\sqrt{1-x^{2}},\\
			\mathcal{G}_{\parallel}^{(3)}(x)&=\frac{3x-x^{3}}{4},\\
			\mathcal{G}_{\perp}^{(3)}(x)&=\frac{3x+x^{3}}{8}.
	\end{align}
	The definition of $\mathcal{G}_{\chi}^{(d)}(x)$ makes it exhibit several nice properties such
	as $\mathcal{G}_{\chi}^{(d)}(-x)=-\mathcal{G}_{\chi}^{(d)}(x)$, $\mathcal{G}_{\chi}^{(d)}(0)=0$, and $\mathcal{G}_{\chi}^{(d)}(\pm1)=\pm1/2$. In addition, we define a more compact notation
	\begin{align}
		\tilde{\zeta}_{\pm}(\omega)=
		\begin{cases}
			+1, &\zeta_{\pm}(\omega)>+1,\\\\
			-1, &\zeta_{\pm}(\omega)<-1,\\\\
			\zeta_{\pm}(\omega), &-1\leq\zeta_{\pm}(\omega)\leq+1.
		\end{cases}
	\end{align}
	In this way, $\Gamma_{\chi}^{(\mathrm{IB})}(\omega,d;\mu, t )$ can be expressed in terms of a unified form
	\begin{align}
		\Gamma_{\chi}^{(\mathrm{IB})}(\omega,d;\mu, t )
		=\mathcal{G}_{\chi}^{(d)}\left[\tilde{\zeta}_{+}(\omega)\right]
		+\mathcal{G}_{\chi}^{(d)}\left[\tilde{\zeta}_{-}(\omega)\right],
	\end{align}
	for all tilted phases ($t=0$, $0<t<1$, $t=1$, and $t>1$) and both components ($\chi=\parallel$ and $\chi=\perp$)
	in 2D and 3D. These unified expressions of $\Gamma_{\chi}^{(\mathrm{IB})}(\omega,d;\mu,t)$
	will be recast more explicitly in the following subsection.
	
	In the regime of large photon energy $\omega=\Omega\gg\omega_+(t)$, one could write
	down the asymptotic background value
	\begin{align}
		\mathrm{Re}\sigma_{\chi}^{ \mathrm{asyp}}(\omega,d)=S^{(\mathrm{IB})}(\Omega,d)\Phi_{\chi}^{(d)}(t),
	\end{align}
	where
	\begin{align}\label{Eq.27}
		\Phi_{\chi}^{(d)}(t)&=\Gamma_{\chi}^{(\mathrm{IB})}(\Omega,d;\mu,t)
		\notag\\&=
		\begin{cases}
			1, & 0\le t \leq1,\\\\
			2\mathcal{G}_{\chi}^{(d)}(1/t)\in(0,1), & t >1,
		\end{cases}
	\end{align}
	which is equal to $\Gamma_{\chi}^{(\mathrm{IB})}(\omega,d;\mu=0,t)$ as well. Especially for $d=2$, $\Phi_\chi^{(d)}(t)$
	is interpreted as the frequency-independent absorption \cite{PRBWild2022}. It is obvious that a different type of tilted Dirac material yields different $\Gamma_{\chi}^{(\mathrm{IB})}(\omega,d;\mu,t)$ and
	hence distinct $\Phi_{\chi}^{(d)}(t)$.
In addition, the dependence of  $\Gamma_{\chi}^{(\mathrm{IB})}(\omega,d;\mu,t)$ and $\Phi_{\chi}^{(d)}(t)$ on the geometry of the Fermi surface in tilted Dirac bands is intuitively illustrated in Appendix \ref{apdx.02}.
	
	\begin{figure*}[htbp]
		\centering
		\includegraphics[width=16cm]{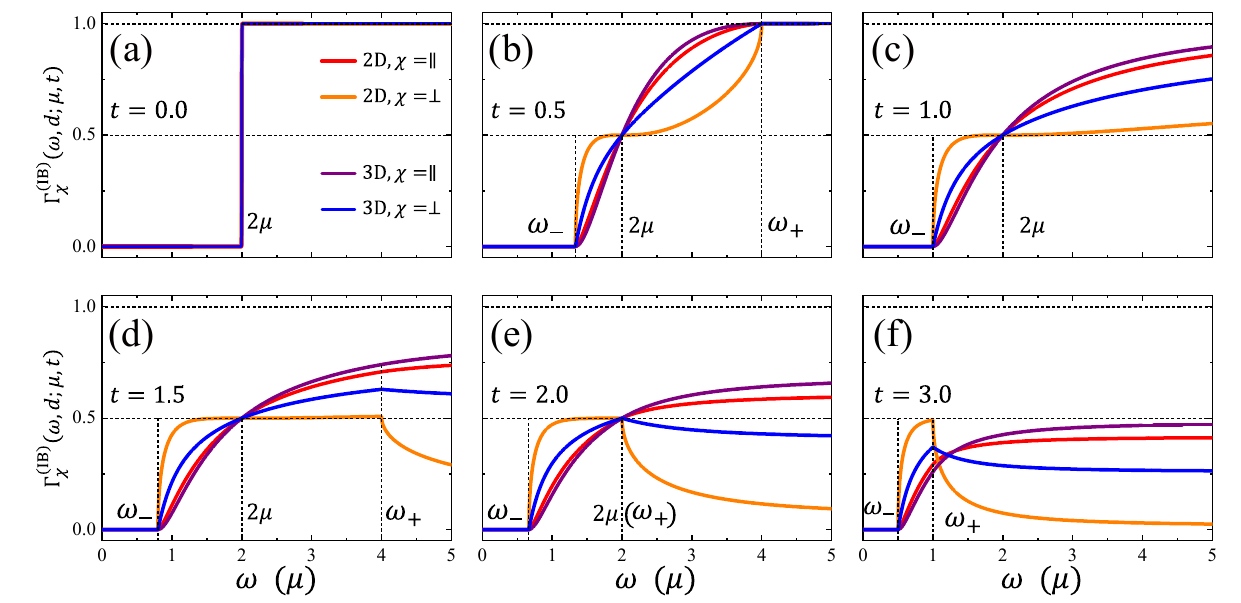}
		\caption{Comparison among $\Gamma_{\parallel}^{(\mathrm{IB})}(\omega,d;\mu,t)$ and $\Gamma_{\perp}^{(\mathrm{IB})}(\omega,d;\mu,t)$ for tilted Dirac bands in 2D and 3D.}
		\label{fig1}
	\end{figure*}
	
	\subsection{Explicit expressions of $\Gamma_{\chi}^{(\mathrm{IB})}(\omega,d;\mu,t)$\label{Sbsec:IB_I}}
	
	The unified expressions of $\Gamma_{\chi}^{(\mathrm{IB})}(\omega,d;\mu,t)$ above can be written more explicitly as follows. For the type-I phase ($0<t<1$), we obtain that
	\begin{align}
		&\Gamma_{\chi}^{(\mathrm{IB})}(\omega,d;\mu,t)
		\notag\\&=
		\begin{cases}
			0, & 0<\omega<\omega_{-}(t),\\\\
			\frac{1}{2}+\mathcal{G}_{\chi}^{(d)}\left[\zeta_{-}(\omega)\right], & \omega_{-}(t)\leq\omega<\omega_{+}(t),\\\\
			1, & \omega\ge\omega_{+}(t).
		\end{cases}
		\nonumber
	\end{align}
	Combined with the expression of $S^{(\mathrm{IB})}(\omega,d)$, these expressions for 2D and 3D can reproduce the results in Refs.\cite{PRBCarbotte2016,PRBJDOS2021,PRBTan2022,PRBTan2021}.
	
	In the untilted limit ($ t \to0^+$), since $\omega_\pm(t)\to2\mu$, the dimensionless functions $\Gamma_{\chi}^{(\mathrm{IB})}(\omega,d;\mu, t )$ restore the previous results for the untilted Dirac cone
	\cite{PRLCarbotte2006,PRLMikhailov2007,PRLMarel2008,PRLMak2008,PRBStauber2008} that $
	\mathop{\lim}\limits_{ t \to 0^+}\Gamma_{\chi}^{(\mathrm{IB})}(\omega,d;\mu, t )
	=\Theta(\omega-2\mu)$.

	For the type-II phase ($ t >1$), we have
	\begin{align}
		&\Gamma_{\chi}^{(\mathrm{IB})}(\omega,d;\mu, t )
		\notag\\&
		=\begin{cases}
			0, & 0<\omega<\omega_{-}(t),\\\\
			\frac{1}{2}+\mathcal{G}_{\chi}^{(d)}\left[\zeta_{-}(\omega)\right], & \omega_{-}(t)\leq\omega<\omega_{+}(t),\\\\
			 \mathcal{G}_{\chi}^{(d)}\left[\zeta_{+}(\omega)\right]+\mathcal{G}_{\chi}^{(d)}\left[\zeta_{-}(\omega)\right], & \omega\geq\omega_{+}(t).
		\end{cases}
		\nonumber
	\end{align}
	Combined with $S^{(\mathrm{IB})}(\omega,d)$ in Table \ref{Tab1}, these expressions of $\Gamma_{\chi}^{(\mathrm{IB})}(\omega,d;\mu,t)$ automatically yield the results for 2D
	and 3D in Refs.\cite{PRBTan2022,PRBCarbotte2016}.
	
	For the type-III phase ($ t =1$), due to the relations $\omega_{-}(t)=\mu$ and $\omega_{+}(t)\to\infty$, we arrive at
	\begin{align}
		\label{Eq22}
		&\Gamma_{\chi}^{(\mathrm{IB})}(\omega,d;\mu, t =1)
		\notag\\&=
		\begin{cases}
			0, & 0<\omega<\mu,\\\\
			\frac{1}{2}+\mathcal{G}_{\chi}^{(d)}\left[\zeta_{-}(\omega)\right], & \omega\geq\mu,
		\end{cases}
	\end{align}
	which can be derived from the limit of both type-I phase and type-II phase. This shows a continuity of $\Gamma_{\chi}^{(\mathrm{IB})}(\omega,d;\mu, t =1)$ and hence interband LOCs, with respect to the tilt parameter, namely,
	\begin{align}
		\lim_{ t \to1^{\pm}}\Gamma_{\chi}^{(\mathrm{IB})}(\omega,d;\mu, t)
		&=\Gamma_{\chi}^{(\mathrm{IB})}(\omega,d;\mu, t =1).
	\end{align}
	Specifically, in 2D these expressions reduce to identical results in our previous work \cite{PRBTan2022}.
	
	From the comparison in Fig. \ref{fig1}, we find some common features of $\Gamma_{\chi}^{(\mathrm{IB})}(\omega,d;\mu,t)$. First, $\Gamma_{\chi}^{(\mathrm{IB})}(\omega,d;\mu,t)$ can be divided into three different regions for both the type-I and type-II phases, but can only be divided into two different regions for
	the type-III phases (see Fig. \ref{fig1}).
	Second, for the untilted phase, $\Gamma_{\chi}^{(\mathrm{IB})}(\omega,d;\mu,t)$ behaves exactly the same, no matter $\chi=\parallel$ or $\chi=\perp$ and $d=2$ or $d=3$. Third, when $0< t\le2$, the dimensionless functions have a universal and robust fixed point
		\begin{align}
			\Gamma_{\chi}^{(\mathrm{IB})}(\omega=2\mu,d;\mu,0<t\le 2)\equiv \frac{1}{2}\Phi_{\chi}^{(d)}(t=0)=\frac{1}{2},
		\end{align}
		no matter $d=2$ or $d=3$ and  $\chi=\perp$ or $\chi=\parallel$. It is noted that $\Phi_{\chi}^{(d)}(t=0)=1$ is nothing but the asymptotic background value of untilted Dirac bands. Although this interesting behavior was previously reported in 2D and 3D tilted Dirac bands \cite{PRBTan2022,PRBCarbotte2016}, the underlying physical reason was lacking therein. Intuitively, the robust behavior can be  understood by the geometric structures of Fermi surface and energy resonance contour (see Appendix \ref{apdx.02} for details).
	Fourth, for tilted phases, $\Gamma_{\perp}^{(\mathrm{IB})}(\omega,3;\mu,t)$ and $\Gamma_{\perp}^{(\mathrm{IB})}(\omega,2;\mu,t)$
	exhibit a distinct difference, in contrast to a close behavior between $\Gamma_{\parallel}^{(\mathrm{IB})}(\omega,3;\mu,t)$ and $\Gamma_{\parallel}^{(\mathrm{IB})}(\omega,2;\mu,t)$. In addition, the difference between $\Gamma_{\parallel}^{(\mathrm{IB})}(\omega,d;\mu,t)$ and $\Gamma_{\perp}^{(\mathrm{IB})}(\omega,d;\mu,t)$ in 3D is not as remarkable as in 2D. Physically, this interesting behavior originates mainly from the increase of the spatial dimension in the measure $\mathrm{d}^d\boldsymbol{k}/(2\pi)^d$ that appears in the definition of LOCs. A larger spatial dimension $d$ dilutes the influence of tilted Dirac dispersion more remarkably. Intuitively, the band tilting of Dirac dispersion is more unremarkable in higher spatial dimensions because more untilted directions in the energy dispersion overwhelmingly suppress the contribution of the tilted direction.
	Combined with the vanished interband LOCs in 1D, we find that 2D can maximize the
	differences in the interband LOCs of the tilted Dirac cone among the spatial dimensions of physical interest. Fifth, the asymptotic background values in the regime of large photon energy ($\omega=\Omega\to\infty$) are identical to
	the results for the undoped case ($\mu=0$).
	
	\section{Intraband conductivity  \label{Sec:Intraband Part}}
	
	The real part of the intraband/Drude LOCs can be written as
	\begin{align}
		\mathrm{Re}\sigma_{jj}^{(\mathrm{D})}(\omega,d)
		&=S_{jj}^{(\mathrm{D})}(d;\mu)\Gamma_{jj}^{(\mathrm{D})}(d;\mu/\Lambda,t)\delta(\omega).\label{Djj}
	\end{align}
	In these notations, the dimensionless function $\Gamma_{jj}^{(\mathrm{D})}(d;\mu/\Lambda,t)$
	characterizes the intraband LOCs for the tilted Dirac cone and $S_{jj}^{(\mathrm{D})}(d;\mu)$ measures
	the scaling of dimensional dependence with respect to chemical potential $\mu$ and spatial dimension $d$, which is explicitly given by
	\begin{align}
		\left[\mathrm{Re}\sigma_{jj}^{(\mathrm{D})}(\omega,d)\right]_{t=0}=S_{jj}^{(\mathrm{D})}(d;\mu)\delta(\omega).
	\end{align}
	It is emphasized that the Drude weight is nothing but $S_{jj}^{(\mathrm{D})}(d;\mu)\Gamma_{jj}^{(\mathrm{D})}(d;\mu/\Lambda,t)$, where $\Lambda$ denotes
	the ultraviolet cutoff scale of momentum proportional to the inverse lattice spacing. The dimension-dependent magnitude
	\begin{align}
		&S_{jj}^{(\mathrm{D})}(d;\mu)
		\notag\\&=
		4\pi g_sg_v\sigma_{0}\sum_{\lambda=\pm}\int_{-\infty}^{+\infty}\frac{\mathrm{d}^{d}\boldsymbol{k}}{(2\pi)^{d}}
		\frac{k_ik_j}{|\boldsymbol{k}|^2}
		\delta\left(\lambda|\boldsymbol{k}|-\mu\right),
	\end{align}
	which can be further expressed as
	\begin{align}
		S_{jj}^{(\mathrm{D})}(d;\mu)
		&=
		\begin{cases}
			S_{\parallel}^{(\mathrm{D})}(d;\mu), & j=1,\\\\
			S_{\perp}^{(\mathrm{D})}(d;\mu), & j\neq 1,
		\end{cases}
	\end{align}
	where
	\begin{align}
		S_{\chi}^{(\mathrm{D})}(d;\mu)\equiv S^{(\mathrm{D})}(d;\mu),
	\end{align}
	with $\chi=\parallel,\perp$. Based on the same reason as in the interband LOCs, the only permitted $S_{\chi}^{(\mathrm{D})}(d;\mu)$ in 1D is $S_{\parallel}^{(\mathrm{D})}(d=1;\mu)\equiv S^{(\mathrm{D})}(d=1;\mu)$
	along the parallel direction, in contrast to the well-defined $S_{\parallel}^{(\mathrm{D})}(d;\mu)$ and $S_{\perp}^{(\mathrm{D})}(d;\mu)$ in 2D and 3D, showing a qualitative difference between 1D and higher dimensions.
	The explicit expressions of $S^{(\mathrm{D})}(d;\mu)$ are listed with respect to $d=1,2,3$ in Table \ref{Tab2}.
	\begin{table}[htbp]
		\begin{tabular*}
			{\columnwidth}{@{\extracolsep{\fill}} c | c c c}
			\hline
			\hline
			$ d$  &$1$ &$2$ &$3$
			\tabularnewline
			\hline
			$S^{(\mathrm{D})}(d;\mu)$  &$16\sigma_{0}$ &$4\sigma_{0}\mu$ & $\frac{1}{v_{F}}\frac{8}{3\pi }\sigma_{0}\mu^{2}$
			\tabularnewline
			\hline
			\hline
		\end{tabular*}
		\caption{Explicit expressions of $S^{(\mathrm{D})}(d;\mu)$ for $d=1,2,3$. Here we restore $v_F$ for explicitness.}
		\label{Tab2}
	\end{table}
	
	From dimensional analysis, one can further obtain a more general result that $S^{(\mathrm{D})}(d;\mu)\propto\sigma_{0}\mu^{d-1}$ for $d\ge1$. In 1D, we find $S^{(\mathrm{D})}(d;\mu)$ is independent of $\mu$. In 2D and 3D, it
	restores the power-law dependence with respect to $\mu$ for the intraband LOCs in Refs. \cite{PRBTan2022,PRBCarbotte2016}. These results also show that 1D is qualitatively different from higher dimensions. Due to this power-law dependence with respect to $\mu$ and particle-hole symmetry, we focus on the $n$-doped case ($\mu>0$) throughout this section.
	
	The dimensionless function $\Gamma_{jj}^{(\mathrm{D})}(d;\mu/\Lambda,t)$ can be categorized to be
	\begin{align}
		\Gamma_{jj}^{(\mathrm{D})}(d;\mu/\Lambda,t)
		&=
		\begin{cases}
			\Gamma_{\parallel}^{(\mathrm{D})}(d;\mu/\Lambda,t), & j=1,\\\\
			\Gamma_{\perp}^{(\mathrm{D})}(d;\mu/\Lambda,t), & j\neq 1,
		\end{cases}
	\end{align}
	for different chemical potential $\mu$, dimension ($d=1,2,3$), and components ($\chi=\parallel$ and $\chi=\perp$). In the untilted phase ($t=0$), the dimensionless function $\Gamma_{\chi}^{(\mathrm{IB})}(\omega,d;\mu, t)=1$.
	
	Consequently, we recast $\mathrm{Re}\sigma_{jj}^{(\mathrm{D})}(\omega,d)$ for the \emph{isotropic} tilted Dirac bands as
	\begin{align}
		\mathrm{Re}\sigma_{jj}^{(\mathrm{D})}(\omega,d)=
		\begin{cases}
			\mathrm{Re}\sigma_{\parallel}^{(\mathrm{D})}(\omega,d), & j=1,\\\\
			\mathrm{Re}\sigma_{\perp}^{(\mathrm{D})}(\omega,d), & j\neq 1,
		\end{cases}
	\end{align}
	where
	\begin{align}
		\mathrm{Re}\sigma_{\chi}^{(\mathrm{D})}(\omega,d)
		=S^{(\mathrm{D})}(d;\mu)\Gamma_{\chi}^{(\mathrm{D})}(d;\mu/\Lambda,t)\delta(\omega)
	\end{align}
	with $d=1,2,3$ and $\chi=\parallel,\perp$. In the following two subsections, we present analytical expressions of $\Gamma_{\chi}^{(\mathrm{D})}(d;\mu/\Lambda,t)$ and the connection between the cutoff $\Lambda$ and the carrier density.
	
	\subsection{Expressions of $\Gamma_{\chi}^{(\mathrm{D})}(d;\mu/\Lambda, t )$}\label{Sbsec:D}
	When $0\le t <1$, due to the closed Fermi surface, the dimensionless functions $\Gamma_{\chi}^{(\mathrm{D})}(d;\mu/\Lambda,t)$ are independent of the cutoff $\Lambda$ and converge when the cutoff $\Lambda$ is set to be infinity. For the type-I phase ($0<t <1$), the dimensionless functions $\Gamma_{\chi}^{(\mathrm{D})}(d;\mu/\Lambda,t)$ are given as
	\begin{align}
		&\Gamma_{\parallel}^{(\mathrm{D})}(d=1;\mu/\Lambda, t )=1,\\
		&\Gamma_{\parallel}^{(\mathrm{D})}(d=2;\mu/\Lambda, t )
		=2\frac{1-\sqrt{1- t ^{2}}}{ t ^{2}},\\
		&\Gamma_{\perp}^{(\mathrm{D})}(d=2;\mu/\Lambda, t )
		=2\frac{1-\sqrt{1- t ^{2}}}{ t ^{2}\sqrt{1- t ^{2}}},\\
		&\Gamma_{\parallel}^{(\mathrm{D})}(d=3;\mu/\Lambda, t )
		=\frac{3}{2 t ^{3}}\left[\ln\frac{1+ t }{1- t }-2 t \right],\\
		&\Gamma_{\perp}^{(\mathrm{D})}(d=3;\mu/\Lambda, t )
		=\frac{3}{4 t ^{3}}\left[\frac{2 t }{1- t ^{2}}-\ln\frac{1+ t }{1- t }\right].
	\end{align}
	The results agree exactly with Refs.\cite{PRBGoerbig2019,PRBCarbotte2016} in 3D and Refs.\cite{PRBHerrera2019,PRBTan2022} in 2D. It can be further verified that
	\begin{align}
		\lim_{ t \to0^+}\Gamma_{\chi}^{(\mathrm{D})}(d;\mu/\Lambda,t)
		&=\Gamma_{\chi}^{(\mathrm{D})}(d;\mu/\Lambda,t=1)=1,
	\end{align}
	consistent with $\Gamma_{\chi}^{(\mathrm{D})}(d;\mu/\Lambda,t=0)=1$ in the untilted phase $(t=0)$.
	
	For the type-II and type-III phases ($t\geq1$), considering the open Fermi surface, one needs to introduce an ultraviolet cutoff scale $\Lambda$. Before going further, we show the general dimensionless function in 1D that
	\begin{align}
		\Gamma_{\parallel}^{(\mathrm{D})}(d=1;\mu/\Lambda,t)&=\frac{|1+t|+|1-t|}{2},
	\end{align}
	which is always independent of the chemical potential $\mu$ and the ultraviolet cutoff scale $\Lambda$. This result also implies a qualitative difference between 1D and higher dimensions.
	
	In the following, we keep the expressions of $\Gamma_{\chi}^{(\mathrm{D})}(d;\mu/\Lambda,t)$ up to the finite term when the ultraviolet cutoff scale $\Lambda$
	is set to be sufficiently large, which in reality is proportional to the inverse lattice spacing. For the type-II phase ($ t >1$), the dimensionless functions
	\begin{align}
		&\Gamma_{\parallel}^{(\mathrm{D})}(d=1;\mu/\Lambda, t )= t ,\\
		&\Gamma_{\parallel}^{(\mathrm{D})}(d=2;\mu/\Lambda, t )
		=\frac{2\sqrt{( t ^{2}-1)^3}}{\pi t }\frac{2\Lambda}{\mu}+\frac{2}{ t ^{2}},\\	
		&\Gamma_{\perp}^{(\mathrm{D})}(d=2;\mu/\Lambda, t )
		=\frac{2\sqrt{ t ^{2}-1}}{\pi t }\frac{2\Lambda}{\mu}-\frac{2}{ t ^{2}},\\
		&\Gamma_{\parallel}^{(\mathrm{D})}(d=3;\mu/\Lambda, t )
		=\frac{3}{2 t ^{3}}\left[\frac{( t ^{2}-1)^{2}}{4}\frac{4\Lambda^{2}}{\mu^{2}}
		\right.\notag\\&\left.\hspace{1.5cm}
		+\ln\frac{4\Lambda^{2}}{\mu^{2}}+\ln\frac{| t ^2-1|}{4}-( t ^2-3)\right],\\
		&\Gamma_{\perp}^{(\mathrm{D})}(d=3;\mu/\Lambda, t )
		=\frac{3}{4 t ^{3}}\left[\frac{ t ^{2}-1}{4}\frac{4\Lambda^{2}}{\mu^{2}}
		\right.\notag\\&\left.\hspace{1.5cm}
		-\ln\frac{4\Lambda^{2}}{\mu^{2}}+\ln|4( t ^2-1)|-\frac{ t ^2-3}{ t ^2-1}\right].
	\end{align}
	Keeping the dominant terms when $\Lambda$ is taken to be sufficiently large, we list the corresponding approximate expressions in Table \ref{Tab3}.
	
	\begin{table}[htbp]
		\begin{tabular*}
			{\columnwidth}{@{\extracolsep{\fill}} c | c c c}
			\hline
			\hline
			$ d$  &$1$ &$2$ &$3$
			\tabularnewline
			\hline
			$\Gamma_{\parallel}^{(\mathrm{D})}(d;\mu/\Lambda, t )$
			&$ t $ &$\frac{\sqrt{( t ^2-1)^3}}{\pi t }\frac{2\Lambda}{\mu}$
			& $\frac{3}{2 t ^3}\left[\frac{( t ^{2}-1)^2}{4}\frac{4\Lambda^{2}}{\mu^{2}}
			+\ln\frac{4\Lambda^{2}}{\mu^{2}}\right]$
			\tabularnewline
			\hline
			$\Gamma_{\perp}^{(\mathrm{D})}(d;\mu/\Lambda, t )$
			&\slash &$\frac{\sqrt{ t ^2-1}}{\pi t }\frac{2\Lambda}{\mu}$
			&$\frac{3}{4 t ^3}\left[\frac{ t ^{2}-1}{4}\frac{4\Lambda^{2}}{\mu^{2}}
			-\ln\frac{4\Lambda^{2}}{\mu^{2}}\right]$
			\tabularnewline
			\hline
			\hline
		\end{tabular*}
		\caption{Approximative expressions of $\Gamma_{\chi}^{(\mathrm{D})}(d;\mu/\Lambda,t)$ for type-II phase ($t>1$) with $d=1,2,3$.}
		\label{Tab3}
	\end{table}

	Two remarks are in order. First, in 2D the results for $\Gamma_{\parallel}^{(\mathrm{D})}
	(d=2;\mu/\Lambda, t )$ and $\Gamma_{\perp}^{(\mathrm{D})}(d=2;\mu/\Lambda, t )$ are the same as that in Ref. \cite{PRBTan2022}, which were not calculated in Ref. \cite{PRBWild2022}. Second, in 3D our result for $\Gamma_{\perp}^{(\mathrm{D})}(d;\mu/\Lambda,t)$ can be confirmed by a direct calculation in spite of disagreeing
	slightly with a previous work \cite{PRBCarbotte2016} which only studied $\Gamma_{\perp}^{(\mathrm{D})}(d;\mu/\Lambda,t)$, leaving $\Gamma_{\parallel}^{(\mathrm{D})}(d;\mu/\Lambda,t)$ unexplored.

	For the type-III phase ($ t =1$), similar results are listed as
	\begin{align}
		&\Gamma_{\parallel}^{(\mathrm{D})}(d=1;\mu/\Lambda, t )=1,\\
		&\Gamma_{\parallel}^{(\mathrm{D})}(d=2;\mu/\Lambda, t )=2,\\
		&\Gamma_{\perp}^{(\mathrm{D})}(d=2;\mu/\Lambda, t )=\frac{4}{\pi}\sqrt{\frac{2\Lambda}{\mu}}-2,\\
		&\Gamma_{\parallel}^{(\mathrm{D})}(d=3;\mu/\Lambda, t )
		=\frac{3}{2}\ln\frac{2\Lambda}{\mu},\\
		&\Gamma_{\perp}^{(\mathrm{D})}(d=3;\mu/\Lambda, t )
		=\frac{3}{8}\frac{2\Lambda}{\mu}+\frac{3}{8}\ln\frac{2\Lambda}{\mu}
		\notag\\&\hspace{3.2cm}
		-\frac{3}{8}(2\ln2+1).
	\end{align}
	Keeping the dominant terms when the cutoff scale $\Lambda$ is sufficiently large, we list the corresponding approximative expressions in Table \ref{Tab4}. It is noted that $\Gamma_{\parallel}^{(\mathrm{D})}(d=2;\mu/\Lambda,t=1)$ and $\Gamma_{\perp}^{(\mathrm{D})}(d=2;\mu/\Lambda,t=1)$ reproduce the results of Ref. \cite{PRBTan2022} for the type-III phase in 2D. $\Gamma_{\parallel}^{(\mathrm{D})}(d;\mu/\Lambda,t=1)$ and $\Gamma_{\perp}^{(\mathrm{D})}(d;\mu/\Lambda,t=1)$ were not calculated in Ref. \cite{PRBWild2022} for 2D and Ref. \cite{PRBCarbotte2016} for 3D.
	
	\begin{table}[htbp]
		\begin{tabular*}
			{\columnwidth}{@{\extracolsep{\fill}} c | c c c}
			\hline
			\hline
			$ d$  &$1$ &$2$ &$3$
			\tabularnewline
			\hline
			$\Gamma_{\parallel}^{(\mathrm{D})}(d;\mu/\Lambda,t)$
			&$1$
			&$2$
			&$\frac{3}{2}\log\frac{2\Lambda}{\mu}$
			\tabularnewline
			\hline
			$\Gamma_{\perp}^{(\mathrm{D})}(d;\mu/\Lambda, t )$
			&\slash &$\frac{4}{\pi}\sqrt{\frac{2\Lambda}{\mu}}$
			& $\frac{3}{8}\frac{2\Lambda}{\mu}+\frac{3}{8}\log\frac{2\Lambda}{\mu}$
			\tabularnewline
			\hline
			\hline
		\end{tabular*}
		\caption{Approximative expressions of $\Gamma_{\chi}^{(\mathrm{D})}(d;\mu/\Lambda,t)$ for type-III phase ($t=1$) with $d=1,2,3$.}
		\label{Tab4}
	\end{table}
	
	\begin{figure*}[htbp]
		\centering
		\includegraphics[width=12cm]{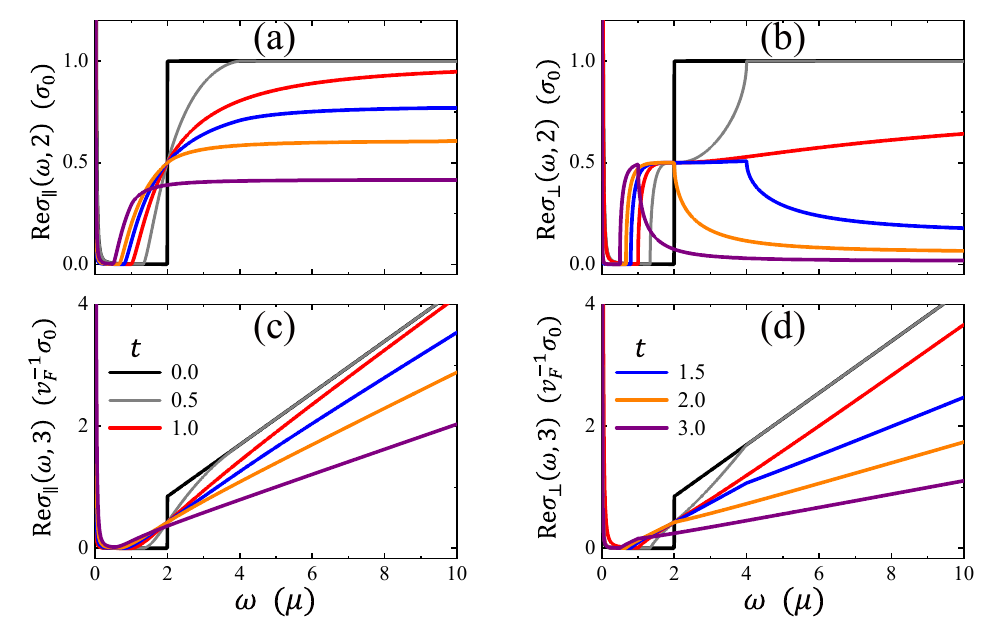}
		\caption{Comparison among $\mathrm{Re}\sigma_{\parallel}(\omega,d)$ and $\mathrm{Re}\sigma_{\perp}(\omega,d)$ for tilted Dirac bands in 2D and 3D.}
		\label{fig2}
	\end{figure*}
	
	Utilizing the analytical results of both interband LOCs and intraband LOCs, we compare $\mathrm{Re}\sigma_{jj}(\omega,d)$ in Fig. \ref{fig2}. The major differences between 2D and 3D are dominant by the scale factor $\sigma_{0}\omega^{d-2}$ in $S_{\chi}^{(\mathrm{IB})}(\omega,d)$, which defines the envelope of interband LOCs for different band tiltings. Accordingly, the impact of band tilting on the LOCs
	in 3D is not as significant as in 2D.
	
	\subsection{Carrier density at small doping}\label{Sbsec:D_DOS}
	
		In this subsection, we derive the carrier density at small doping in $d$ dimension, which is defined as
		\begin{align}
			n(t,d,\mu,\Lambda)=\sum_{\kappa=\pm}\sum_{\lambda=\pm}n^{\lambda}_{\kappa}(t,d,\mu,\Lambda),\label{Eq36}
		\end{align}
		where $n^{\lambda}_{\kappa}(t,d,\mu,\Lambda)$ denotes the carrier density at the $\kappa=\pm$ valley and the $\lambda=\pm$ band. At zero temperature,
		\begin{align}
			n^{\lambda}_{\kappa}(t,d,\mu,\Lambda)&=\lim_{T\rightarrow0^{+}}
			g_{s} \sum_{\boldsymbol{k}}^{|\boldsymbol{k}|\leq\Lambda}
			f\left[\varepsilon^{\lambda}_{\kappa}(\boldsymbol{k})\right]
			\notag\\&=
			g_{s}\mathop{\int}\limits _{|\boldsymbol{k}|\leq\Lambda}
			\frac{\mathrm{d}^{d}\boldsymbol{k}}{(2\pi)^{d}}
			\tilde\Theta\left[\mu-\varepsilon^{\lambda}_{\kappa}(\boldsymbol{k})\right],
			\label{Eq38}
		\end{align}
		which is definitely determined by the carriers below Fermi energy. Since the intraband LOC in 1D or type-I phase is cutoff-independent, it is unnecessary to calculate the relation between carrier density and ultraviolet cutoff; hence we restrict our analysis to $t\geq1$ and $d\geq 2$ hereafter.

		For type-II phase ($t>1$) and type-III phase ($t=1$), an ultraviolet cutoff $\Lambda<+\infty$ is needed in calculating the carrier density due to open Fermi surface, which will dramatically influence the carrier density via the Heaviside function in Eq.(\ref{Eq38}). For type-II phase in 2D, the conduction band ($\lambda=+$) is related to one branch of hyperbolic Fermi surface satisfying $\mu\ge\varepsilon_{\kappa}^{+}(\boldsymbol{k})$, while the valence band ($\lambda=-$) contributes to the other branch of hyperbolic Fermi surface satisfying $\mu\ge\varepsilon_{\kappa}^{-}(\boldsymbol{k})$. It is convenient to introduce two critical cutoffs
		\begin{align}
			\Lambda_{-\lambda}=\frac{\lambda \mu}{1+\lambda t}
		\end{align}
		for conduction band ($\lambda=+$) and valence band ($\lambda=-$), respectively. By contrast, for type-III phase in 2D, the conduction band ($\lambda=+$) is related to the parabolic Fermi surface $\mu\ge\varepsilon_{\kappa}^{+}(\boldsymbol{k})$, while the valence band ($\lambda=-$) does not contribute to the Fermi surface. It is stressed for 3D tilted Dirac bands that the Fermi surfaces are the ellipsoid, paraboloid, and hyperboloid by rotating the ellipse, parabola, and hyperbola (Fermi surfaces in 2D tilted Dirac bands) with respect to the tilting direction, respectively.
	
		The total carrier density for type-II and type-III phases ($t\ge 1$) takes the following explicit expressions as (see Appendix \ref{AppxC} for details)
		\begin{widetext}
			\begin{align}
				n(t\ge 1,d,\mu,\Lambda)&
				=n_{0}(d,\Lambda)
				-\Theta\left(\Lambda-\Lambda_{-}\right)\delta n_{+}(t,d,\mu,\Lambda)
				-\Theta\left(\Lambda-\Lambda_{+}\right)\delta n_{-}(t,d,\mu,\Lambda)
				\nonumber\\&
				=n_{0}(d,\Lambda)
				-\Theta\left[t-\frac{\mu-\Lambda}{\Lambda}\right]
				\delta n_{+}(t,d,\mu,\Lambda)
				-\Theta\left[t-\frac{\mu+\Lambda}{\Lambda}\right]
				\delta n_{-}(t,d,\mu,\Lambda),
			\end{align}
			where $n_{0}(d,\Lambda)$ denotes the mutual contribution from both conduction band ($\lambda=+$) and
			valence band ($\lambda=-$), which satisfies $n_{0}(d,\Lambda)=\frac{2\Lambda^{2}}{\pi}$ for $d=2$ and $n_{0}(d,\Lambda)=\frac{4\Lambda^{3}}{3\pi^{2}}$ for $d=3$, and
			\begin{align}
				\delta n_{\lambda}(t,d,\mu,\Lambda)=\begin{cases}
					\frac{\Lambda^{2}}{\pi^{2}}\arccos\frac{\mu-\lambda\Lambda}{t\Lambda}
					+\frac{\mu^{2}}{\pi^{2}}\mathcal{I}_{\text{2D}}(\frac{\mu-\lambda\Lambda}{t\Lambda},\lambda t), & d=2,\\
					\\
					\frac{\Lambda^{3}}{3\pi^{2}}\left(1-\frac{\mu-\lambda\Lambda}{t\Lambda}\right)
					+\lambda\frac{\mu^{3}}{3\pi^{2}}\mathcal{I}_{\text{3D}}(\frac{\mu-\lambda\Lambda}{t\Lambda},\lambda t), & d=3
				\end{cases}
			\end{align}
			represents the additional contribution from either conduction band ($\lambda=+$) or valence band ($\lambda=-$). Note that two auxiliary functions are introduced as
			\begin{align}
				\mathcal{I}_{\text{2D}}(x,t)=
				\begin{cases} \frac{1}{t^{2}-1}\left[\frac{t\sqrt{1-x^{2}}}{1+tx}-\frac{2}{\sqrt{t^{2}-1}}
					\mathrm{artanh}\sqrt{\frac{(t-1)(1-x)}{(t+1)(1+x)}}\right], & t>1,\\
					\\
					\frac{x+2}{3(x+1)^{2}}\sqrt{1-x^{2}}, & t=1,
				\end{cases}
			\end{align}
			and
			\begin{align}
				\mathcal{I}_{\text{3D}}(x,t)=\frac{1}{2t}\left[\frac{1}{(tx+1)^{2}}-\frac{1}{(t+1)^{2}}\right],
			\end{align}
			which satisfy that $\mathcal{I}_{\mathrm{2D}}(x,t)\to \mathcal{I}_{\mathrm{2D}}(x,1)$ and $\mathcal{I}_{\mathrm{3D}}(x,t)\to \mathcal{I}_{\mathrm{3D}}(x,1)$ in the limit $t\to1^{+}$.
		\end{widetext}

		For $\Lambda<\Lambda_{+}$ or $1< t<\frac{\mu+\Lambda}{\Lambda}$, the additional term $\delta n_{-}(t,d,\mu,\Lambda)$ does not contribute to the total carrier density due to $\Theta\left(\Lambda-\Lambda_{+}\right)\to0$. It is consequently found that
		\begin{align}
			\lim_{t\to 1^{+}}n(1< t<\frac{\mu+\Lambda}{\Lambda},d,\mu,\Lambda)=n(t=1,d,\mu,\Lambda),
		\end{align}
		which indicates that $n(t=1,d,\mu,\Lambda)$ can be consistently obtained by taking the limit $t\to 1^{+}$ of $n(t>1,d,\mu,\Lambda)$ when the cutoff $\Lambda$ is less than $\Lambda_{+}$. This suggests that the total carrier density for type-II phase can smoothly approach that for type-III phase.

		For $\Lambda>\Lambda_{+}$ or $ t>\frac{\mu+\Lambda}{\Lambda}>1$, the total carrier density for type-III phase $n(t=1,d,\mu,\Lambda)$ can not be obtained by taking the limit $t\to 1^{+}$ of the total carrier density for type-II phase $n(t>1,d,\mu,\Lambda)$, namely,
		\begin{align}
			\lim_{t\to 1^{+}}n(t>\frac{\mu+\Lambda}{\Lambda}>1,d,\mu,\Lambda)\neq n(t=1,d,\mu,\Lambda),
		\end{align}
		because the additional term $\delta n_{-}(t,d,\mu,\Lambda)$ also contributes to the total carrier density. Furthermore, if $\Lambda\gg\Lambda_{+}$, we can expand $n(t,d,\mu,\Lambda)$ in the series of $\mu/\Lambda$.
		Up to the leading order of $\mu$, we have
		\begin{align}
			n(t>1,d,\mu,\Lambda)=\begin{cases}
				\frac{\Lambda^{2}}{\pi}+\frac{\Lambda\mu}{\pi^{2}\sqrt{t^{2}-1}}, & d=2,\\
				\\
				\frac{2\Lambda^{3}}{3\pi^{2}}+\frac{\Lambda^{2}\mu}{\pi^{2}t}, & d=3,
			\end{cases}
		\end{align}
		for the type-II phase ($t>1$), and
		\begin{align}
			n(t=1,d,\mu,\Lambda)=\begin{cases}
				\frac{\Lambda^{2}}{\pi}+\frac{2\Lambda}{3\pi^{2}}\sqrt{2\mu\Lambda}, & d=2,\\
				\\
				\frac{2\Lambda^{3}}{3\pi^{2}}+\frac{\Lambda^{2}\mu}{3\pi^{2}}, & d=3,
			\end{cases}
		\end{align}
		for the type-III phase ($t=1$).

		After introducing
		\begin{align}
			\Delta n(t,d,\mu,\Lambda)=n(t,d,\mu,\Lambda)-n(t,d,0,\Lambda),
		\end{align}
		one can extract the information about $\Lambda$ from $n(t,d,\mu,\Lambda)$ at small doping ($\mu\to0$) through
		\begin{align}
			\frac{\Delta n(t>1,d,\mu,\Lambda)}{\mu}&=\mathcal{C}(t>1,d)\Lambda^{d-1},
			\label{CarrDenOvertilted}
		\end{align}
		and
		\begin{align}
			\frac{\Delta n(t=1,d,\mu,\Lambda)}{\mu^{\frac{d-1}{2}}}&=\mathcal{C}(t=1,d)\Lambda^{\frac{d+1}{2}},
			\label{CarrDenCriticaltilted}
		\end{align}
		where two coefficients $\mathcal{C}(t>1,d)$ and $\mathcal{C}(t=1,d)$ can be read out from Eqs.(\ref{CarrDenOvertilted}) and (\ref{CarrDenCriticaltilted}) as
		\begin{align}
			\mathcal{C}(t>1,d)=\begin{cases}
				\frac{1}{\pi^{2}\sqrt{t^{2}-1}}, & d=2,\\
				\\
				\frac{1}{\pi^{2}t}, & d=3,
			\end{cases}
			\label{CCOvertilted}
		\end{align}
		and
		\begin{align}
			\mathcal{C}(t=1,d)=\begin{cases}
				\frac{2\sqrt{2}}{3\pi^{2}}, & d=2,\\
				\\
				\frac{1}{3\pi^{2}}, & d=3.
			\end{cases}
		\end{align}	
		Particularly for $t>1$, the density of states (DOS) near the Dirac point ($\mu=0$) can be written as
		\begin{align}
			N(t,d,0)&=\left.\frac{\partial n(t,d,\mu,\Lambda)}{\partial\mu}\right|_{\mu=0}
			=\lim_{\mu\to0}\frac{\Delta n(t,d,\mu,\Lambda)}{\mu}\notag\\
			&=\sum_{\kappa\pm}\sum_{\lambda=\pm}g_{s}\mathop{\int}\limits _{|\boldsymbol{k}|\leq\Lambda}\frac{\mathrm{d}^{d}\boldsymbol{k}}{(2\pi)^{d}}\delta\left[-\varepsilon^{\lambda}_{\kappa}(\boldsymbol{k})\right]\label{Eq39}.
		\end{align}
		By setting $d=3$ and $t>1$, it yields the same result $N(t,3,0,\Lambda)=\Lambda^{2}/(\pi^2 t)$ as that in Refs. \cite{PRBCarbotte2016,ZyuzinandTiwari} up to a factor of degeneracy. Hence the intraband LOCs are strongly related to the geometric structure of the Fermi surface, via the relation between carrier density and DOS. For example, when $t>1$, the intraband LOC $\mathrm{Re}\sigma_{jj}^{(\mathrm{D})}(\omega,d)$ takes
		\begin{align}
			\mathrm{Re}\sigma_{jj}^{(\mathrm{D})}(\omega,d)
			=\begin{cases}
				 S_{jj}^{(\mathrm{D})}(2;\mu)\Gamma_{jj}^{(\mathrm{D})}(2;\frac{\mu}{\Lambda_{\text{2D}}},t)\delta(\omega), & d=2,\\
				\\
				 S_{jj}^{(\mathrm{D})}(3;\mu)\Gamma_{jj}^{(\mathrm{D})}(3;\frac{\mu}{\Lambda_{\text{3D}}},t)\delta(\omega), & d=3,
			\end{cases}
			\nonumber
		\end{align}
		with $\Lambda_{\text{2D}}=\pi^{2}\sqrt{t^{2}-1}N(t,2,0)$ and $\Lambda_{\text{3D}}=\pi \sqrt{t} \left[N(t,3,0)\right]^{1/2}$.

	\begin{figure*}[htbp]
		\centering
		\includegraphics[width=18cm]{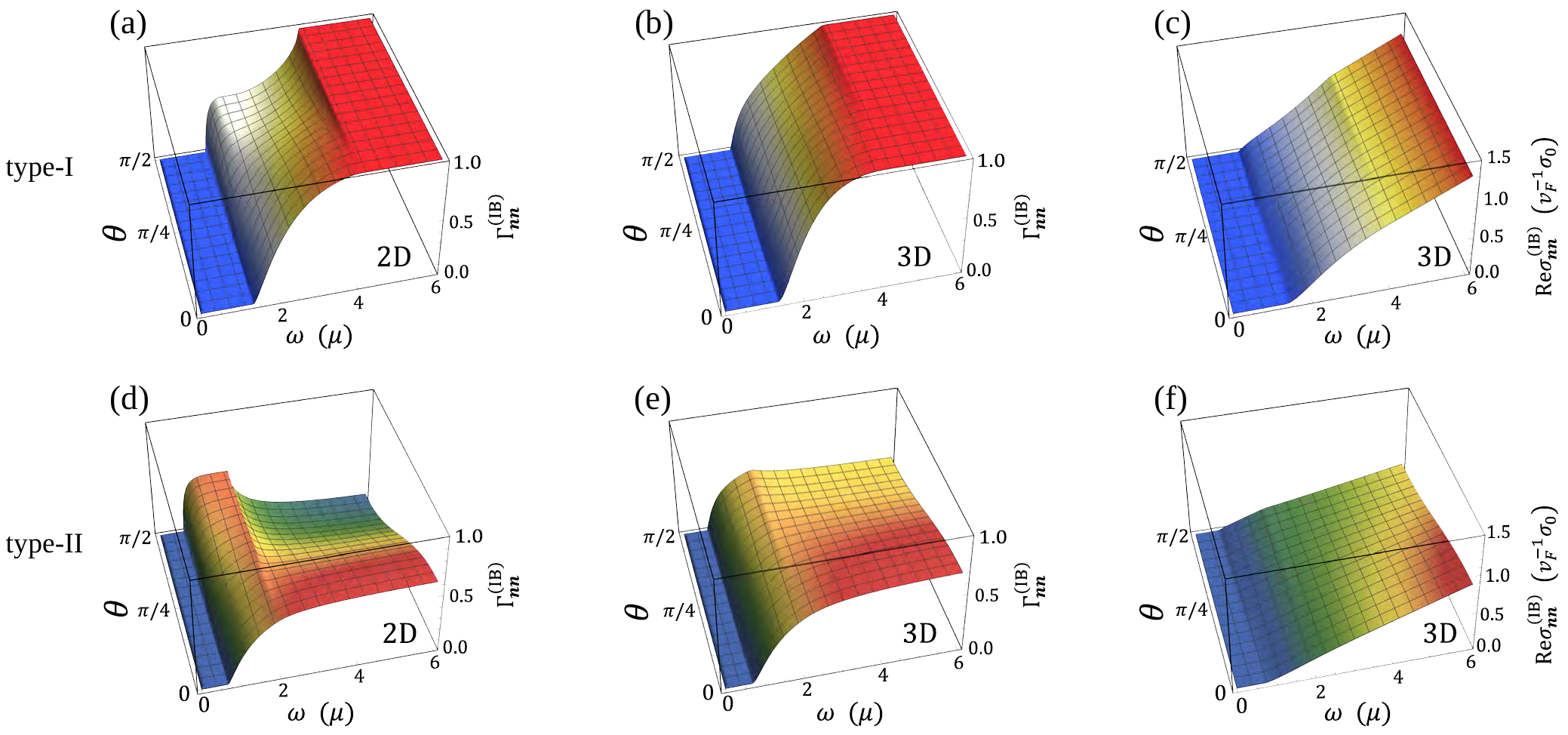}
		\caption{Comparison among interband LOCs for tilted Dirac bands in 2D and 3D. The upper frame and lower frame
			are for the type-I phase (with $t=0.5$) and type-II phase (with $t=2.0$), respectively. Panels (a) and (d) denote $\Gamma_{\boldsymbol{nn}}^{(\mathrm{IB})}(\omega,2;\mu,t)=\mathrm{Re}\sigma_{\boldsymbol{nn}}^{(\mathrm{IB})}
			(\omega,2)/\sigma_0$ in 2D. Panels (b) and (e) represent $\Gamma_{\boldsymbol{nn}}^{(\mathrm{IB})}(\omega,3;\mu,t)$
			in 3D and panels (c) and (f) show the corresponding interband LOCs $\mathrm{Re}\sigma_{\boldsymbol{nn}}^{(\mathrm{IB})}(\omega,3)/(v_{F}^{-1}\sigma_0)$ with $\mathrm{Re}\sigma_{\boldsymbol{nn}}^{(\mathrm{IB})}(\omega,3)=\frac{2\sigma_{0}\omega}{3\pi v_{F}}
			\Gamma_{\boldsymbol{nn}}^{(\mathrm{IB})}(\omega,3;\mu,t)$. Here we restore $v_F$ for explicitness.}
		\label{fig3}
	\end{figure*}

	\section{Angular dependence\label{Sec:Angular dependence}}
	
	The results above are focused on the LOCs along the direction either parallel or perpendicular to the tilting. To present the angular dependence, we choose an arbitrary direction $\boldsymbol{n}=(n_{1},\cdots,n_{d})$ with $|\boldsymbol{n}|=1$. The corresponding LOC can be written as \cite{PRBJDOS2021}
	\begin{align}
		\label{Eq44}
		&\mathrm{Re}\sigma_{\boldsymbol{nn}}(\omega,d)
		\notag\\&
		=\mathrm{Re}\sigma_{\parallel}(\omega,d)\cos^{2}\theta+\mathrm{Re}\sigma_{\perp}(\omega,d)\sin^{2}\theta,
	\end{align}
	where $\theta$ measures the angle between the detection direction $\boldsymbol{n}$ and the tilting direction.
	
	From the definition above, the Drude conductivities read
	\begin{align}
		\mathrm{Re}\sigma_{\boldsymbol{nn}}^{(\mathrm{D})}(\omega,d)
		&=S^{(\mathrm{D})}(d;\mu)\Gamma_{\boldsymbol{nn}}^{(\mathrm{D})}(d;\mu/\Lambda,t)~\delta(\omega),
	\end{align}
	where
	\begin{align}
		\Gamma_{\boldsymbol{nn}}^{(\mathrm{D})}(d;\mu/\Lambda,t)&=
		A^{(\mathrm{D})}(0,d)+B^{(\mathrm{D})}(0,d)\cos^2\theta,
	\end{align}
	with
	\begin{align}
		A^{(\mathrm{D})}(0,d)&=\Gamma_{\perp}^{(\mathrm{D})}(d;\mu/\Lambda, t ),\notag\\
		B^{(\mathrm{D})}(0,d)&=\Gamma_{\parallel}^{(\mathrm{D})}(d;\mu/\Lambda, t )
		-\Gamma_{\perp}^{(\mathrm{D})}(d;\mu/\Lambda, t ).\notag
	\end{align}
	
	For the untilted phase ($t=0$), using the relations $\Gamma_{\parallel}^{(\mathrm{D})}
	(d;\mu/\Lambda,t )=\Gamma_{\perp}^{(\mathrm{D})}(d;\mu/\Lambda,t)=1$, one obtains
	\begin{align}
		\mathrm{Re}\sigma_{\boldsymbol{nn}}^{(\mathrm{D})}(\omega,d)|_{t=0}&=S^{(\mathrm{D})}(d;\mu)\delta(\omega),
	\end{align}
	which indicates a direction-independent Drude peak. However, for the tilted phases ($t>0$), due to the relation $\Gamma_{\parallel}^{(\mathrm{D})}
	(d;\mu/\Lambda,t)\neq\Gamma_{\perp}^{(\mathrm{D})}(d;\mu/\Lambda,t)$, the term $B^{(\mathrm{D})}(0,d)$ does not vanish due to the tilted Dirac band. This allows us to propose an approach to determine whether the Dirac band is tilted or not, by measuring the angular dependence of the Drude peak.
	
	The real part of the interband LOCs reads
	\begin{align}
		\mathrm{Re}\sigma_{\boldsymbol{nn}}^{(\mathrm{IB})}(\omega,d)
		&=S^{(\mathrm{IB})}(\omega,d)\Gamma_{\boldsymbol{nn}}^{(\mathrm{IB})}(\omega,d;\mu,t),
	\end{align}
	where
	\begin{align}
		\Gamma_{\boldsymbol{nn}}^{(\mathrm{IB})}(\omega,d;\mu,t)
		&=A^{(\mathrm{IB})}(\omega,d)+B^{(\mathrm{IB})}(\omega,d)\cos^{2}\theta,
	\end{align}
	with
	\begin{align}
		A^{(\mathrm{IB})}(\omega,d)&=\Gamma_{\perp}^{(\mathrm{IB})}(\omega,d;\mu,t),\notag\\
		B^{(\mathrm{IB})}(\omega,d)&=\Gamma_{\parallel}^{(\mathrm{IB})}(\omega,d;\mu,t)
		-\Gamma_{\perp}^{(\mathrm{IB})}(\omega,d;\mu,t).\notag
	\end{align}
	
	The comparison among the angular dependence of interband LOCs for tilted Dirac bands in 2D and 3D is shown in Fig. \ref{fig3}. Note that $\Gamma_{\boldsymbol{n}\boldsymbol{n}}^{(\mathrm{IB})}(\omega,d;\mu,t)$ in 2D is sharper than that in 3D. The boundaries of $\Gamma_{\boldsymbol{n}\boldsymbol{n}}^{(\mathrm{IB})}(\omega,d;\mu,t)$ at $\theta=0$ and $\theta=\frac{\pi}{2}$ reduce to $\Gamma_{\parallel}^{(\mathrm{IB})}(\omega,d;\mu,t)$ and $\Gamma_{\perp}^{(\mathrm{IB})}(\omega,d;\mu,t)$
	shown in Fig. \ref{fig1}(b) and Fig. \ref{fig1}(e), respectively. It is emphasized that in Figs. \ref{fig3}(a) and \ref{fig3}(d), the boundaries at $\theta=0$ and $\theta=\frac{\pi}{2}$ give rise to the corresponding $\mathrm{Re}\sigma_{\parallel}^{(\mathrm{IB})}(\omega,2)/\sigma_0$ in Fig. \ref{fig2}(a) and $\mathrm{Re}\sigma_{\perp}^{(\mathrm{IB})}(\omega,2)/\sigma_0$ in Fig. \ref{fig2}(b).
	Besides, $\mathrm{Re}\sigma_{\parallel}^{(\mathrm{IB})}(\omega,3)/(v_F^{-1}\sigma_0)$ in Fig. \ref{fig2}(c) and $\mathrm{Re}\sigma_{\perp}^{(\mathrm{IB})}(\omega,3)/(v_F^{-1}\sigma_0)$ in Fig. \ref{fig2}(d) are nothing but the
	boundaries at $\theta=0$ and $\theta=\frac{\pi}{2}$ shown in Figs. \ref{fig3}(c) and \ref{fig3}(f).
	
	In the regime of large photon energy ($\omega=\Omega\to\infty$), the asymptotic background value of the interband LOCs is
	\begin{align}
		&\mathrm{Re}\sigma_{\boldsymbol{nn}}^{\mathrm{asyp}}(\omega,d)
		\notag\\&=
		S^{(\mathrm{IB})}(\Omega,d)\left[A^{(\mathrm{IB})}(\Omega,d)+B^{(\mathrm{IB})}(\Omega,d)\cos^{2}\theta\right],
	\end{align}
	where
	\begin{align}
		A^{(\mathrm{IB})}(\Omega,d)&=\Phi_{\perp}(t),\notag\\
		B^{(\mathrm{IB})}(\Omega,d)&=\Phi_{\parallel}(t)-\Phi_{\perp}(t).\notag
	\end{align}
	
	For cases with $t\le1$, using $\Phi_{\parallel}(t)=\Phi_{\perp}(t)=1$, one obtains
	\begin{align}
		\mathrm{Re}\sigma_{\boldsymbol{nn}}^{\mathrm{asyp}}(\omega,d)&=S^{(\mathrm{IB})}(\Omega,d),
	\end{align}
	which indicates a directional independence with respect to the detection direction in the asymptotic background value of the interband LOC. However, for type-II phase ($t>1$), one has $\Phi_{\parallel}(t)\neq\Phi_{\perp}(t)$, which implies $B^{(\mathrm{IB})}(\Omega,d)\neq 0$ due to the over-tilting of type-II Dirac band. This way, one may determine whether $t$ is greater than unity by measuring the angular dependence of the asymptotic background value of interband LOCs.
	
	\section{Effect of anisotropy \label{Sec:Effect of anisotropy}}
	
	The above analysis was restricted to the \emph{isotropic} tilted Dirac cones. To show the effect of anisotropy on the LOCs, we turn to the Hamiltonian of \emph{anisotropic} tilted Dirac fermions in $d$ dimension
	\begin{align}
		\mathcal{H}_{\kappa}(\boldsymbol{k})=\kappa\hbar v_{t}k_{1}\tau_{0}+\sum_{j=1}^{d}\hbar v_{j}k_{j}\tau_{j},
	\end{align}
	where the spatial dimension $d\geq 2$ and the \emph{anisotropic} Fermi velocities $v_j$ with $j=1,2,\cdots$ are not necessarily equal to the \emph{isotropic} Fermi velocity $v_F$.
	
	\begin{table}[htbp]
		\begin{tabular*}
			{\columnwidth}{@{\extracolsep{\fill}} c | c c}
			\hline
			\hline
			$ d$  &$2$ &$3$
			\tabularnewline
			\hline
			$S^{(\mathrm{IB})}(\omega,d)$ & $\sigma_{0}$ & $\frac{1}{v_{F}}\frac{2}{3\pi}\sigma_{0}\omega$
			\tabularnewline
			\hline
			$S_{jj}^{(\mathrm{IB})}(\omega,d)$ & $\frac{v_{j}^2}{v_{1}v_{2}}\sigma_{0}$
			& $\frac{v_{j}^2}{v_{1}v_{2}v_{3}}\frac{2}{3\pi}\sigma_{0}\omega$
			\tabularnewline
			\hline
			$S^{(\mathrm{D})}(d;\mu)$
			&$4\sigma_{0}\mu$
			& $\frac{1}{v_{F}}\frac{8}{3\pi}\sigma_{0}\mu^{2}$
			\tabularnewline
			\hline
			$S_{jj}^{(\mathrm{D})}(d;\mu)$&
			$\frac{v_{j}^2}{v_{1}v_{2}}4 \sigma_{0} \mu $ &
			$\frac{v_{j}^2}{v_{1}v_{2}v_{3}}\frac{8}{3\pi}\sigma_{0}\mu^{2}$
			\\
			\hline
			\hline
		\end{tabular*}
		\caption{The $jj$-component of LOCs for \emph{anisotropic} and \emph{isotropic} tilted Dirac cones in 2D and 3D. Here we restore $v_F$ for explicitness.}
		\label{Tab5}
	\end{table}
	
	Detailed calculations show that the decomposition in Eqs. (\ref{IBjj}) and (\ref{Djj}) still hold and that the forms of $\Gamma_{jj}^{(\mathrm{IB})}(\omega,d;\mu,t)$
	and $\Gamma_{jj}^{(\mathrm{D})}(d;\mu/\Lambda,t)$ do not change. However, $S_{jj}^{(\mathrm{IB})}(\omega,d)$ and $S_{jj}^{(\mathrm{D})}(d;\mu)$ can not be written as $S_{\chi}^{(\mathrm{IB})}(\omega,d)$ and $S_{\chi}^{(\mathrm{D})}(d;\mu)$ any more. The essential effect of anisotropy can be generally expressed in terms of the ratio consisting of Fermi velocities as
	\begin{align}
		S_{jj}^{(\mathrm{IB})}(\omega,d)
		&=\frac{v_j^2}{\Pi_{i=1}^{d}v_i}S^{(\mathrm{IB})}(\omega,d),\\
		S_{jj}^{(\mathrm{D})}(d;\mu)
		&=\frac{v_j^2}{\Pi_{i=1}^{d}v_i}S^{(\mathrm{D})}(d;\mu),
	\end{align}	
	where $S^{(\mathrm{IB})}(\omega,d)$ and $S^{(\mathrm{D})}(d;\mu)$ denote the counterparts in the \emph{isotropic} model with $v_j=v_F$. Interestingly, the tilt parameter $t$ does not enter the ratio in the above relations. Specifically, for the untilted phase ($t=0$) and type-I phase ($0<t<1$), the $jj$-component of LOCs for \emph{anisotropic}
	and \emph{isotropic} tilted Dirac cones in 2D and 3D are explicitly tabulated in Table \ref{Tab5}, which agrees with the results in Refs. \cite{PRBVerma2017,PRBTan2021,PRBHerrera2019,PRBWild2022,PRBTan2022}.
	
	\section{SUMMARY \label{Sec:Summary}}
	
	In this work, we formulated a unified theory for analyzing LOC in general terms of the spatial dimensionality (1D-3D) of materials with tilted Dirac cones. Depending on the tilt parameter $t$, the Dirac electrons have four phases: untilted, type-I, type-II, and type-III; the Dirac dispersion can also be \emph{isotropic} and/or \emph{anisotropic}. As such the LOC has multitudes of possible behaviors and a unified theory is very useful. Amongst the multitudes of situations, there were individual analyses in the literature and our unified theory reproduces to these known results for the corresponding situations. More importantly, our comparable study not only reports qualitative differences/similarities of different spatial dimensionality and band tilting, but also reveals several robust properties independent of spatial dimensionality.

	It was found that for 1D tilted Dirac bands, $\mathrm{Re}\sigma_{jj}^{(\mathrm{IB})}(\omega,d=1)=0$
	and $\mathrm{Re}\sigma_{jj}^{(\mathrm{D})}(\omega,d=1)=8\left(|1+t|+|1-t|\right)\sigma_{0}\delta(\omega)$. From dimensional analysis, the scaling behaviors $\mathrm{Re}\sigma_{jj}^{(\mathrm{IB})}(\omega,d)\propto S_{jj}^{(\mathrm{IB})}(\omega,d)\propto\sigma_{0}
		\omega^{d-2}$ and $\mathrm{Re}\sigma_{jj}^{(\mathrm{D})}(\omega,d)\propto S_{jj}^{(\mathrm{D})}(\omega,d)\propto\sigma_{0}\mu^{d-1}$ always hold, which dominates the major differences between the LOCs in 2D and 3D. These results indicate that 1D is qualitatively different from higher dimensions. For tilted phases, $\Gamma_{\chi}^{(\mathrm{IB})}(\omega,d;\mu,t)$ shows qualitative similarities among $d=2$ and $d=3$, $\chi=\perp$ and $\chi=\parallel$. In addition, the quantitative difference between $\Gamma_{\parallel}^{(\mathrm{IB})}(\omega,d;\mu,t)$ and $\Gamma_{\perp}^{(\mathrm{IB})}(\omega,d;\mu,t)$ in 3D is not as significant as in 2D, originating mainly from the fact that the larger spatial dimensionality of integration measure dilutes the influence of tilted Dirac dispersion more remarkably. When $0<t\le2$, the universal and robust behavior $\Gamma_{\chi}^{(\mathrm{IB})}(\omega=2\mu,d;\mu,t)\equiv 1/2$ holds no matter $d=2$ or $d=3$, $\chi=\perp$ or $\chi=\parallel$, which can be intuitively understood by the geometric structures of Fermi surface and energy resonance contour. The intraband LOCs and the carrier density for 2D and 3D tilted Dirac bands are both closely related to the geometric structure of the Fermi surface and the cutoff of integration. The angular dependence of LOCs can be used to characterize both spatial dimensionality and band tilting and the constant asymptotic background value reflects the feature of Dirac bands. The LOCs in the \emph{anisotropic} tilted Dirac cone can be connected to its \emph{isotropic} counterpart by a ratio that consists of Fermi velocities, no matter in 2D or 3D.

	Most of our findings are universal for tilted Driac bands and hence valid for a great many Dirac materials (untilted, type-I,-II,and -III), in the spatial dimension (1D - 3D) of physical interest \cite{RMP2009,RMP2018,SatoPRB2018,Yangnjp2020,Zhunjp2020,WangACSNano2021,Zhangnjp2022,YueNanoLett2022,JPSJ2006,Zhou8Pmmn2014PRL,Science8Pmmn2015,Nature2015,Volovik2017,Volovik2018,Liu2021}, comprising several cases of Dirac nanoribbon (1D and untitled), graphene (2D and untilted), $\alpha \text{-(BEDT-TTF)}_{2}\text{I}_{3}$ (2D and type-I), and 8\emph{Pmmn}-borophene in the presence of vertical electric field (2D and type-I, -II, and -III). Theoretical predictions of this work are experimentally testable through the spectromicroscopic study \cite{ZQLi2008,PRLMak2008}.

	\section*{ACKNOWLEDGEMENTS\label{Sec:acknowledgements}}
	
	
	This work is partially supported by the National Natural Science Foundation of China under Grants No. 11547200, No. 11874271, and No. 11874273. J.-T.H. acknowledges financial support from Sichuan Science and Technology under Grant No.2022-YCG057. We also gratefully acknowledge financial support from NSERC of Canada (H.R.C. and H.G.) and the FQRNT of the Province of Quebec (H.G.).
	
	\appendix
	\begin{widetext}

		\section{Basic definitions and Major formulas of longitudinal optical conductivities}\label{apdx.01}
		The longitudinal optical conductivities ($\omega\geq0$) are given by definition as
		\begin{align}
			\sigma_{jj}^{\kappa}(\omega,d)=\frac{\mathrm{i}}{\omega}k_{B}Te^{2}
			\lim_{\boldsymbol{q}\to\boldsymbol{0}}
			 \int_{-\infty}^{\infty}\frac{\mathrm{d}^d\boldsymbol{k}}{(2\pi)^{d}}\sum_{\Omega_m}\mathrm{Tr}\left[\frac{\partial H_{\kappa}(\boldsymbol{k})}{\partial k_{j}}G_{\kappa}(\boldsymbol{k},\mathrm{i}\Omega_{m})\frac{\partial H_{\kappa}(\boldsymbol{k})}{\partial k_{j}}G_{\kappa}(\boldsymbol{k}+\boldsymbol{q},\mathrm{i}\Omega_{m}+\omega+\text{i}\eta)\right],
		\end{align}
		where the spatial index $j=1,2...d$ and the Matsubara Green's function
		$ G_{\kappa}(\boldsymbol{k},\text{i}\Omega_{m})=[(\text{i}\Omega_{m}+\mu)\tau_{0}-H_{\kappa}(\boldsymbol{k})]^{-1}
		$ with
		$\Omega_{m}=\frac{2\pi k_{B}T}{\hbar}(m+\frac{1}{2})$.
		Following the same procedure of our previous work \cite{PRBMojarro2022,PRBTan2021} and setting $\hbar=k_B=v_F=1$,
		it is easy to obtain
		\begin{align}
			\sigma_{jj}^{\kappa}(\omega,d)=\frac{\text{i}}{\omega}\lim_{\boldsymbol{q}\to\boldsymbol{0}}
			\int_{-\infty}^{+\infty}\frac{\mathrm{d}^{d}\boldsymbol{k}}{(2\pi)^{d}}\sum_{\lambda,\lambda^{\prime}=\pm}
			\mathcal{F}_{\lambda,\lambda^{\prime}}^{\kappa;jj}(\boldsymbol{k},\boldsymbol{k}+\boldsymbol{q})
			 \frac{f[\varepsilon^{\lambda}_{\kappa}(\boldsymbol{k})]-f[\varepsilon_{\kappa}^{\lambda^{\prime}}(\boldsymbol{k}+\boldsymbol{q})]}{\omega+\varepsilon^{\lambda}_{\kappa}(\boldsymbol{k})-\varepsilon_{\kappa}^{\lambda^{\prime}}(\boldsymbol{k}+\boldsymbol{q})+\text{i}\eta},
		\end{align}
		where
		\begin{align}
			\mathcal{F}_{\lambda,\lambda^{\prime}}^{\kappa;jj}(\boldsymbol{k},\boldsymbol{k}+\boldsymbol{q})&=
			\frac{e^{2}}{2}\left\{ t^{2}\delta_{j1}\delta_{j1}\left[1+\lambda\lambda'\frac{\boldsymbol{k}\cdot(\boldsymbol{k}+\boldsymbol{q})}{|\boldsymbol{k}||\boldsymbol{k}+\boldsymbol{q}|}\right]
			+2\kappa t
			 \delta_{j1}\left[\lambda\frac{k_{j}}{|\boldsymbol{k}|}+\lambda'\frac{k_{j}+q_{j}}{|\boldsymbol{k}+\boldsymbol{q}|}\right]\right\}
			\nonumber\\&
			 +\frac{e^{2}}{2}\left\{\delta_{jj}\left[1-\lambda\lambda'\frac{\boldsymbol{k}\cdot(\boldsymbol{k}+\boldsymbol{q})}{|\boldsymbol{k}||\boldsymbol{k}+\boldsymbol{q}|}\right]+2\lambda\lambda'\frac{k_{j}(k_{j}+q_{j})}{|\boldsymbol{k}||\boldsymbol{k}+\boldsymbol{q}|}\right\}.
		\end{align}
		
		By utilizing the Dirac identity $\frac{1}{x+i\eta}=\mathcal{P}\frac{1}{x}-i\pi\delta(x)$, one recast the real part as
		\begin{align}
			\text{Re}\sigma_{jj}^{\kappa}(\omega,d)
			= & -\text{i}^{2}\pi\sum_{\lambda,\lambda^{\prime}=\pm}
			\lim_{\boldsymbol{q}\to\boldsymbol{0}}
			\int_{-\infty}^{\infty}\frac{\text{d}^{d}\boldsymbol{k}}{(2\pi)^{d}}
			 \mathcal{F}_{\lambda,\lambda'}^{\kappa;jj}(\boldsymbol{k},\boldsymbol{k}+\boldsymbol{q})\frac{f\left[\varepsilon^{\lambda}_{\kappa}(\boldsymbol{k})\right]
				-f\left[\varepsilon_{\kappa}^{\lambda^{\prime}}(\boldsymbol{k}+\boldsymbol{q})\right]}{\omega}
			\delta\left[\omega+\varepsilon^{\lambda}_{\kappa}(\boldsymbol{k})
			-\varepsilon_{\kappa}^{\lambda^{\prime}}(\boldsymbol{k}+\boldsymbol{q})\right]
			\nonumber\\
			= & -\pi\sum_{\lambda,\lambda^{\prime}=\pm}
			\lim_{\boldsymbol{q}\to\boldsymbol{0}}
			\int_{-\infty}^{\infty}\frac{\text{d}^{d}\boldsymbol{k}}{(2\pi)^{d}}
			\mathcal{F}_{\lambda,\lambda'}^{\kappa;jj}(\boldsymbol{k},\boldsymbol{k}+\boldsymbol{q})
			\frac{f\left[\varepsilon^{\lambda}_{\kappa}(\boldsymbol{k})\right]
				 -f\left[\varepsilon_{\kappa}^{\lambda^{\prime}}(\boldsymbol{k}+\boldsymbol{q})\right]}{\varepsilon^{\lambda}_{\kappa}(\boldsymbol{k})
				 -\varepsilon_{\kappa}^{\lambda^{\prime}}(\boldsymbol{k}+\boldsymbol{q})}\delta\left[\omega+\varepsilon^{\lambda}_{\kappa}(\boldsymbol{k})
			-\varepsilon_{\kappa}^{\lambda^{\prime}}(\boldsymbol{k}+\boldsymbol{q})\right].
		\end{align}
		where $\omega$ in the denominator
		was substituted by $-\left[\varepsilon^{\lambda}_{\kappa}(\boldsymbol{k})
		-\varepsilon_{\kappa}^{\lambda^{\prime}}(\boldsymbol{k}+\boldsymbol{q})\right]$ due to the appearance of $\delta\left\{\omega+\left[\varepsilon^{\lambda}_{\kappa}(\boldsymbol{k})
		-\varepsilon_{\kappa}^{\lambda^{\prime}}(\boldsymbol{k}+\boldsymbol{q})\right]\right\}$ in the integration.
		
		Since the summation over band indices $\lambda$ and $\lambda'$ can be decomposed into $\sum_{\lambda,\lambda'=\pm}=\sum_{\lambda\lambda'=-}+\sum_{\lambda\lambda'=+}$, the corresponding real part of LOC at a given valley $\kappa$ can be divided into an interband part and an intraband (Drude) part as
		\begin{align}
			\mathrm{Re}\sigma_{jj}^{\kappa}(\omega,d)=\mathrm{Re}\sigma_{jj}^{\kappa(\mathrm{IB})}(\omega,d)
			+\mathrm{Re}\sigma_{jj}^{\kappa(\mathrm{D})}(\omega,d),
		\end{align}
		where
		\begin{align}
			\text{Re}\sigma_{jj}^{\kappa(\text{IB})}(\omega,d)= & \pi\sum_{\lambda=\pm}\sum_{\lambda^{\prime}=-\lambda} \lim_{\boldsymbol{q}\to\boldsymbol{0}}\int_{-\infty}^{\infty}\frac{\text{d}^{d}\boldsymbol{k}}{(2\pi)^{d}}
			\mathcal{F}_{\lambda,\lambda'}^{\kappa;jj}(\boldsymbol{k},\boldsymbol{k}+\boldsymbol{q})
			\frac{f\left[\varepsilon^{\lambda}_{\kappa}(\boldsymbol{k})\right]
				-f\left[\varepsilon_{\kappa}^{\lambda^{\prime}}(\boldsymbol{k}+\boldsymbol{q})\right]}{\omega}
			\delta\left[\omega+\varepsilon^{\lambda}_{\kappa}(\boldsymbol{k})
			-\varepsilon_{\kappa}^{\lambda^{\prime}}(\boldsymbol{k}+\boldsymbol{q})\right]
			\nonumber\\
			= & \pi\sum_{\lambda=\pm}\int_{-\infty}^{\infty}\frac{\text{d}^{d}\boldsymbol{k}}{(2\pi)^{d}}
			 \mathcal{F}_{\lambda,-\lambda}^{\kappa;jj}(\boldsymbol{k})\frac{f\left[\varepsilon^{\lambda}_{\kappa}(\boldsymbol{k})\right]
				-f\left[\varepsilon_{\kappa}^{-\lambda}(\boldsymbol{k})\right]}{\omega}
			 \delta\left[\omega+\varepsilon^{\lambda}_{\kappa}(\boldsymbol{k})-\varepsilon_{\kappa}^{-\lambda}(\boldsymbol{k})\right]
			\nonumber\\
			=& \pi\int_{-\infty}^{\infty}\frac{\text{d}^{d}\boldsymbol{k}}{(2\pi)^{d}}
			\mathcal{F}_{-,+}^{\kappa;jj}(\boldsymbol{k})\frac{f\left[\varepsilon_{\kappa}^{-}(\boldsymbol{k})\right]
				-f\left[\varepsilon_{\kappa}^{+}(\boldsymbol{k})\right]}{\omega}
			\delta\left[\omega+\varepsilon_{\kappa}^{-}(\boldsymbol{k})
			-\varepsilon_{\kappa}^{+}(\boldsymbol{k})\right]
			\nonumber\\
			= & \pi\int_{-\infty}^{\infty}\frac{\text{d}^{d}\boldsymbol{k}}{(2\pi)^{d}}
			\mathcal{F}_{-,+}^{\kappa;jj}(\boldsymbol{k})\frac{f\left[\varepsilon_{\kappa}^{-}(\boldsymbol{k})\right]
				 -f\left[\varepsilon_{\kappa}^{+}(\boldsymbol{k})\right]}{\omega}\delta\left(\omega-2|\boldsymbol{k}|\right)
			\nonumber\\
			=& 4\pi\sigma_{0}\int_{-\infty}^{+\infty}\frac{\mathrm{d}^{d}\boldsymbol{k}}{(2\pi)^{d}}
			\left[\delta_{jj}-\frac{k_jk_j}{|\boldsymbol{k}|^2}\right]
			\frac{f\left[\varepsilon_{\kappa}^{-}(\boldsymbol{k})\right]
				-f\left[\varepsilon_{\kappa}^{+}(\boldsymbol{k})\right]}{\omega}
			\delta\left(\omega-2|\boldsymbol{k}|\right),
		\end{align}
		and
		\begin{align}
			\text{Re}\sigma_{jj}^{\kappa(\text{D})}(\omega,d)= & -\pi\sum_{\lambda=\pm}\sum_{\lambda^{\prime}=\lambda}
			\lim_{\boldsymbol{q}\to\boldsymbol{0}}\int_{-\infty}^{\infty}\frac{\text{d}^{d}\boldsymbol{k}}{(2\pi)^{d}}
			\mathcal{F}_{\lambda,\lambda'}^{\kappa;jj}(\boldsymbol{k},\boldsymbol{k}+\boldsymbol{q})
			\frac{f\left[\varepsilon^{\lambda}_{\kappa}(\boldsymbol{k})\right]
				-f\left[\varepsilon_{\kappa}^{\lambda^{\prime}}(\boldsymbol{k}+\boldsymbol{q})\right]}{
				\varepsilon^{\lambda}_{\kappa}(\boldsymbol{k})
				-\varepsilon_{\kappa}^{\lambda^{\prime}}(\boldsymbol{k}+\boldsymbol{q})}
			\delta\left[\omega+\varepsilon^{\lambda}_{\kappa}(\boldsymbol{k})
			-\varepsilon_{\kappa}^{\lambda^{\prime}}(\boldsymbol{k}+\boldsymbol{q})\right]
			\nonumber\\
			= & -\pi\sum_{\lambda=\pm}\lim_{\boldsymbol{q}\to\boldsymbol{0}}
			\int_{-\infty}^{\infty}\frac{\text{d}^{d}\boldsymbol{k}}{(2\pi)^{d}}
			\mathcal{F}_{\lambda,\lambda}^{\kappa;jj}(\boldsymbol{k},\boldsymbol{k}+\boldsymbol{q})
			\frac{f\left[\varepsilon^{\lambda}_{\kappa}(\boldsymbol{k})\right]
				 -f\left[\varepsilon^{\lambda}_{\kappa}(\boldsymbol{k}+\boldsymbol{q})\right]}{\varepsilon^{\lambda}_{\kappa}(\boldsymbol{k})-\varepsilon^{\lambda}_{\kappa}(\boldsymbol{k}+\boldsymbol{q})}
			\delta\left[\omega+\varepsilon^{\lambda}_{\kappa}(\boldsymbol{k})
			-\varepsilon_{\kappa}^{\lambda}(\boldsymbol{k}+\boldsymbol{q})\right]
			\nonumber\\
			=&
			\pi\sum_{\lambda=\pm}\int_{-\infty}^{+\infty}\frac{\mathrm{d}^{d}\boldsymbol{k}}{(2\pi)^{d}}
			\mathcal{F}_{\lambda,\lambda}^{\kappa;jj}(\boldsymbol{k})
			 \left\{-\frac{\mathrm{d}f\left[\varepsilon^{\lambda}_{\kappa}(\boldsymbol{k})\right]}{\mathrm{d}\varepsilon^{\lambda}_{\kappa}(\boldsymbol{k})}\right\}
			\delta(\omega)
			\nonumber\\
			=&4\pi\sigma_{0}\sum_{\lambda=\pm}\int_{-\infty}^{+\infty}\frac{\mathrm{d}^{d}\boldsymbol{k}}{(2\pi)^{d}}
			\frac{\left(\kappa t |\boldsymbol{k}|\delta_{j1}+\lambda k_j\right)
				\left(\kappa t |\boldsymbol{k}|\delta_{j1}+\lambda k_j\right)}{|\boldsymbol{k}|^2}			 \left\{-\frac{\mathrm{d}f\left[\varepsilon^{\lambda}_{\kappa}(\boldsymbol{k})\right]}{\mathrm{d}\varepsilon^{\lambda}_{\kappa}(\boldsymbol{k})}\right\}
			\delta(\omega),
		\end{align}
		with $\mu\ge0$ assumed and
		\begin{align}
			\mathcal{F}_{\lambda,\lambda^{\prime}}^{\kappa;jj}(\boldsymbol{k})\equiv\lim_{\boldsymbol{q}\to 0}	 \mathcal{F}_{\lambda,\lambda^{\prime}}^{\kappa;jj}(\boldsymbol{k},\boldsymbol{k}+\boldsymbol{q})
			 =e^{2}(1-\delta_{\lambda\lambda^{\prime}})\left[\delta_{jj}-\frac{k_{j}k_{j}}{|\boldsymbol{k}|^{2}}\right]+e^{2}\delta_{\lambda\lambda^{\prime}}\frac{\left(\kappa t|\boldsymbol{k}|\delta_{j1}+\lambda k_{j}\right)\left(\kappa t|\boldsymbol{k}|\delta_{j1}+\lambda k_{j}\right)}{|\boldsymbol{k}|^{2}}.
		\end{align}
		It is noted that the final three equations are nothing but Eqs. (\ref{Eq.7}) -(\ref{Eq.9}) in the main text.

		\section{Intuitive analysis for the behaviors of $\Gamma_{jj}^{\text{(IB)}}(\omega,d;\mu,t)$}\label{apdx.02}

		This appendix aims to present an intuitive understanding for the behaviors of $\Gamma_{jj}^{\text{(IB)}}(\omega,d;\mu,t)$ from the geometric structure of Fermi surface
		and energy resonance contour. For later convenience, we define two critical cutoffs
		\begin{align}
			\Lambda_{\pm}=\frac{\mu}{|1\mp t|},
		\end{align}
		which can also be written as $\frac{\omega_{\pm}(t)}{2}$ in the main text.
		
		In the presence of optical field with frequency $\omega>0$, the interband LOC at $\omega$ is generally
		determined by the optical transition governed by $\left\{f\left[\varepsilon_\kappa^{-}(\boldsymbol{k})\right] -f\left[\varepsilon_\kappa^{+}(\boldsymbol{k})\right]\right\}
		\delta\left[\omega+\varepsilon_\kappa^{-}(\boldsymbol{k})-\varepsilon_\kappa^{+}(\boldsymbol{k})\right]$, where $\left\{f\left[\varepsilon_\kappa^{-}(\boldsymbol{k})\right] -f\left[\varepsilon_\kappa^{+}(\boldsymbol{k})\right]\right\}$ and $\delta\left[\omega+\varepsilon_\kappa^{-}(\boldsymbol{k})-\varepsilon_\kappa^{+}(\boldsymbol{k})\right]$ are
		imposed by Pauli principle and energy conservation, respectively. Concretely,
		the energy resonance contour is a sphere with radius $r(d)=|\boldsymbol{k}|=\omega/2$ in $d$-dimensional wavevector space. In addition, for the $n$-doped case ($\mu>0$), the corresponding
		Fermi surfaces can be obtained from $\varepsilon^{\lambda}_{\kappa}(\boldsymbol{k})=\kappa t  k_{1}
		+\lambda|\boldsymbol{k}|=\mu$.
		
		\begin{figure*}[h]
			\centering
			\includegraphics[width=15cm]{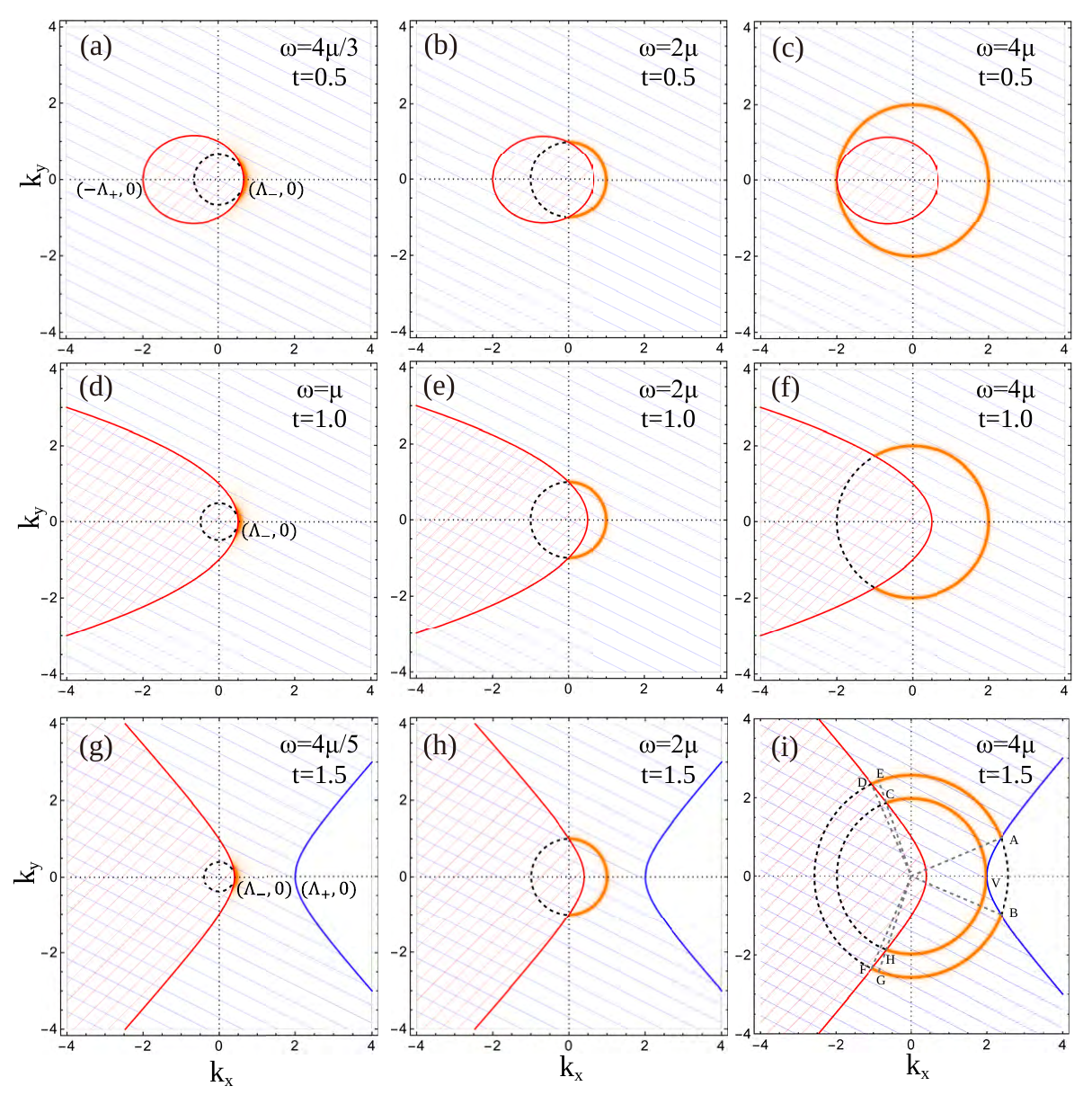}
			\caption{Fermi surfaces and energy resonance contours with (a)-(c) for type-I phase, (d)-(f) for type-III phase,
				and (g)-(i) for type-II phase. Due to Pauli principle, the dashed black arcs represent that the corresponding interband optical transition are forbidden, whereas the orange stripe arcs represent that the corresponding interband optical transition are allowed.}
			\label{FigAppxB1}
		\end{figure*}
		
		In the following, we present our explicit analysis for the 2D tilted Dirac bands at first, and then emphasize how to apply this analysis to the 3D case.
		In 2D, the Fermi surface projected to the $k_x$-$k_y$ plane are conic sections where the tilt parameter $t$ is nothing
		but the eccentricity. As shown in Fig.\ref{FigAppxB1}, the elliptic ($0<t<1$) and parabolic ($t=1$) Fermi surfaces
		are represented by red lines for conduction band, and the hyperbolic ($t>1$) Fermi surfaces are represented by red
		line and blue line for conduction band and valence band, respectively. The $k_x$-$k_y$ plane can be divided into two kinds of region termed ``Allowed Region'' ($\mathscr{R}_{a}$) and ``Forbidden Region'' ($\mathscr{R}_{f}$), due to Pauli principle. Concretely, the region between two branches of hyperbolic Fermi surfaces or the region outside the elliptic/parabolic Fermi surfaces represents the fact that the valence band below the Fermi surface is occupied but the conduction band is unoccupied, which allows the interband optical transition due to Pauli principle, hence termed as the ``Allowed Region'' $\mathscr{R}_{a}$ (denoted by diagonal light blue lines). By contrast, the region inside elliptic Fermi surfaces or the region on the left side of parabolic/hyperbolic Fermi surfaces denotes that the conduction band below the Fermi surface and the valence band below the conduction band are both occupied, which forbids the interband optical transition due to Pauli principle, hence termed as the ``Forbidden Region'' $\mathscr{R}_{f1}$ (denoted by meshed red and blue lines). Besides, the region on the right side of the hyperbolic Fermi surface corresponds to the fact that both the conduction band below the Fermi surface and the valence band above the Fermi surface are unoccupied, which forbids the interband optical transition due to Pauli principle, hence termed the ``Forbidden Region'' $\mathscr{R}_{f2}$ (denoted by nothing/white). The energy resonance contours are denoted by a circle with radius $r(d=2)=|\boldsymbol{k}|=\omega/2$ shown in Fig.\ref{FigAppxB1}. The dashed black arc(s) and orange stripe arc(s) correspond to the energy resonance contours in the ``Forbidden Region'' and ``Allowed Region'', respectively. The states involved in the interband optical transition are gathered around orange stripe arc(s). Keeping these in mind, we intuitively analyze the behaviors of $\Gamma_{jj}^{\text{(IB)}}(\omega,d;\mu,t)$ as follows, which measures the percentage of carriers contributing to interband optical transition. Intuitively, one can geometrically extract the information of  $\Gamma_{jj}^{\text{(IB)}}(\omega,d;\mu,t)$ from  $ (\mathrm{the~central~angle~of~orange~stripe~arcs})/2\pi$.

		For the type-I phase ($0<t<1$), when the radius $r(d=2)\le \Lambda_{-}$ [see Fig.\ref{FigAppxB1}(a)], the interband optical transition is forbidden by Pauli principle. Intuitively, the central angle of orange stripe arcs is zero, hence $\Gamma_{jj}^{\text{(IB)}}(\omega,d=2;\mu,t)\equiv 0$. When the radius $r(d=2)\ge \Lambda_{+}$ [see Fig.\ref{FigAppxB1}(c)], the interband optical
		transition can always take place no matter how large the radius is. Intuitively, the central angle of orange stripe arcs is $2\pi$, which results in the fact that the percentage of contributing carriers keep invariant $\Gamma_{jj}^{\text{(IB)}}(\omega,d=2;\mu,t)\equiv 1$, denoting the asymptotic background
		value. When the radius $r(d=2)$ increases from $\Lambda_{-}$ to $\Lambda_{+}$ [see Fig.\ref{FigAppxB1}(b)], the corresponding central angle of orange stripe arcs increases from 0 to $2\pi$ monotonously. The percentage of carriers contributing to the interband optical transition increases from 0 to 1 in this region.
		
		For the type-III phase ($t=1$), when the radius $r(d=2)\le \Lambda_{-}$ [see Fig.\ref{FigAppxB1}(d)], the interband optical transition can not occur due to Pauli blocking, hence $\Gamma_{jj}^{\text{(IB)}}(\omega,d=2;\mu,t)\equiv 0$. However, the percentage of carriers contributing to the interband optical transition (or intuitively the central angle of orange stripe arcs) is monotonously increasing with radius $r(d=2)>\Lambda_{-}$, as shown in Figs.\ref{FigAppxB1}(e) and (f). It is noted that the percentage is always less than 1 even when the radius $r(d=2)\to +\infty$. Consequently, when $r(d=2)>\Lambda_{-}$, one always has $0<\Gamma_{jj}^{\text{(IB)}}(\omega,d=2;\mu,t)<1$.
		
		\begin{figure*}[h]
			\centering
			\includegraphics[width=18cm]{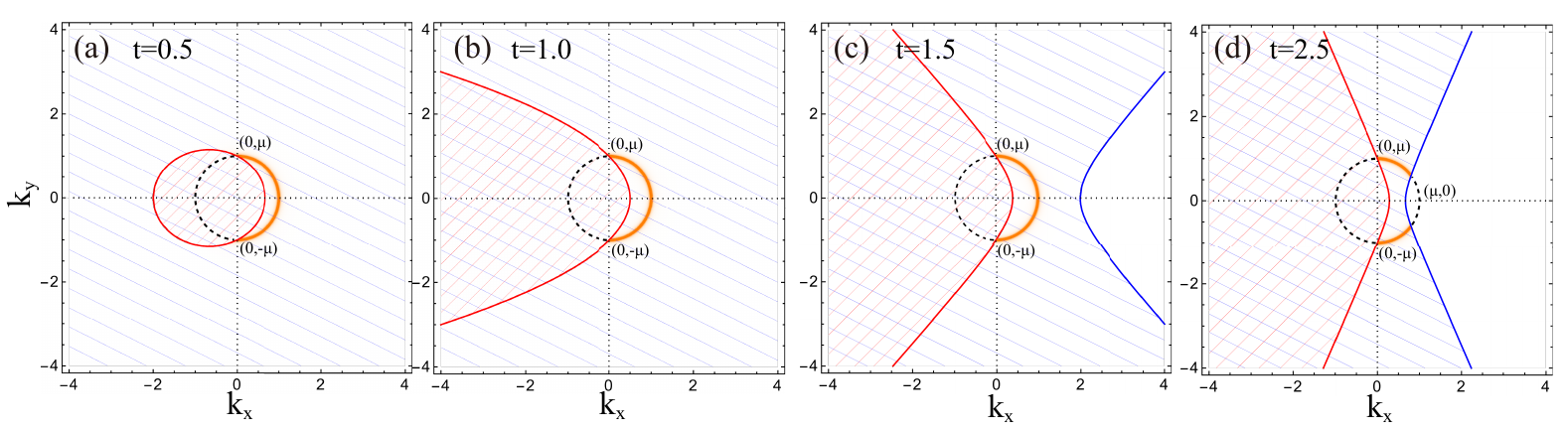}
			\caption{Universal behavior of fixed point at $\omega=2\mu$ for $0<t\le 2$ in (a)-(c), while no such fixed point at $\omega=2\mu$ in (d).}
			\label{figA2-02}
		\end{figure*}
		
		For the type-II phase ($t>1$), when the radius $r(d=2)\le \Lambda_{-}$ [see Fig.\ref{FigAppxB1}(g)], the interband optical transition is not allowed, hence $\Gamma_{jj}^{\text{(IB)}}(\omega,d=2;\mu,t)\equiv 0$.
		When the radius $\Lambda_{-}<r(d=2)\le\Lambda_{+}$, the percentage of carriers contributing to the interband optical
		transition increases with the radius [see Fig.\ref{FigAppxB1}(h)]. When the radius $r(d=2)>\Lambda_{+}$, the behavior
		of $\Gamma_{jj}^{\text{(IB)}}(\omega,d=2;\mu,t)$ can also be analyzed by taking into account the complexities brought
		by the competition between its two distinct increments shown in Fig.\ref{FigAppxB1}(i): one dashed arc
		(denoted by $\overset{\frown}{AB}$) causing a nonpositive contribution and two orange stripe arcs (denoted by $\overset{\frown}{DE}$ and $\overset{\frown}{FG}$) causing a positive contribution.

		We now turn to the universal behavior of fixed point $\Gamma_{jj}^{\text{(IB)}}(\omega=2\mu,d=2;\mu,0< t\le2)=1/2$
		at $\omega=2\mu$. From Fig. \ref{figA2-02} (a-c), when $0< t\le 2$, the whole energy resonance contour with radius $r(d=2)=\mu$ can always be divided into an orange stripe semi-circle and a dashed black semi-circle. Intuitively, the central angle of orange stripe arcs is always $\pi$, hence the percentage of carriers contributing to the interband optical transition always accounts for $1/2$, namely, $\Gamma_{jj}^{\text{(IB)}}(\omega=2\mu,d=2;\mu,0< t\le2)=1/2$. By contrast, from Fig. \ref{figA2-02} (d), there is dashed line in the ``Forbidden Region'' on the right side of hyperbolic Fermi surfaces marked in blue, hence the
		orange stripe arc can account only for the percentage less than $1/2$. Therefore, $\Gamma_{\chi}^{(\mathrm{IB})}(\omega=2\mu,\mu,t>2)$ is naturally smaller than that for $0<t<2$, namely, $\Gamma_{jj}^{\text{(IB)}}(\omega=2\mu,d=2;\mu,t>2)<1/2$.
		
		\begin{figure*}[h]
			\centering
			\includegraphics[width=13.5cm]{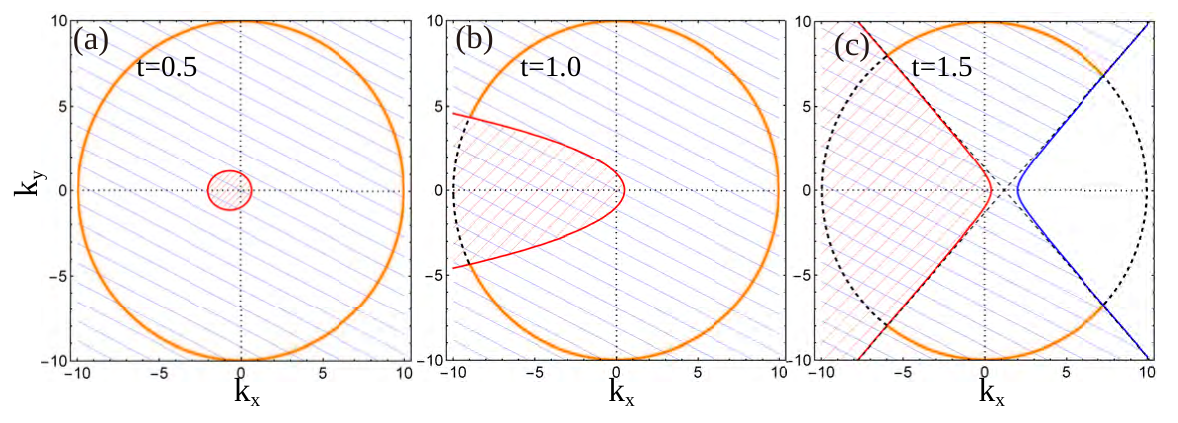}
			\caption{Intuitive understanding of asymptotic background values of $\Gamma_{jj}^{\text{(IB)}}(\omega=\Omega\to+\infty,d=2;\mu,t)$.}
			\label{figAAppxB}
		\end{figure*}
		
		It is time to present an intuitive analysis of asymptotic background values of $\Gamma_{jj}^{\text{(IB)}}(\omega,d=2;\mu,t)$ at $\omega=\Omega\to +\infty$. From Fig. \ref{figAAppxB}(a),
		when $0<t<1$, the interband optical
		transition can always take place no matter how large the radius is. Intuitively, the central angle of orange stripe arcs is always $2\pi$, which results in the fact that the asymptotic background
		value keep invariant $\Gamma_{jj}^{\text{(IB)}}(\omega=\Omega\to+\infty,d=2;\mu,t)\equiv 1$. From Fig. \ref{figAAppxB}(b), when $t=1$, in the asymptotic region $\omega=\Omega\to+\infty$, the central angle of orange stripe arcs is less than but approaching $2\pi$, hence the percentage of contributing carriers is always less than but approaching 1. From Fig. \ref{figAAppxB}(c), when $t>1$, the central angle of orange stripe arcs is uniquely determined by the slope of asymptote of hyperbola $\mathscr{K}_{\text{asym}}=\pm\sqrt{t^2-1}$, hence there is a one-to-one mapping between $\Phi_{\parallel}^{(2)}(t)+\Phi_{\perp}^{(2)}(t)$ and the eccentricity of hyperbola (tilting parameter $t$), namely,
		\begin{align}
			\Phi_{\parallel}^{(2)}(t)+\Phi_{\perp}^{(2)}(t)=\frac{4}{\pi}\arcsin\left(\frac{1}{t}\right)= \frac{4}{\pi}\mathrm{arccot}\left(|\mathscr{K}_{\text{asym}}|\right).
		\end{align}
		
		It is emphasized that the essential spirit of the intuitive analysis above can apply to 3D tilted Dirac bands where the Fermi surfaces are the ellipsoid, paraboloid, and hyperboloid by rotating the ellipse, parabola, and hyperbola (Fermi surfaces in 2D tilted Dirac bands) with respect to the tilting direction, respectively. Due to the axial symmetry with respect to tilting direction, the Fermi surface and energy resonance contour for 3D tilted Dirac bands can reduce to that for 2D case presented above.
		
		\section{Relation between carrier density and ultraviolet cutoff \label{AppxC}}

		In this appendix, we show the detailed derivation of the relation between carrier density and ultraviolet cutoff for type-II phase ($t>1$) and type-III phase ($t=1$). Hereafter, we restrict our analysis to $t\geq1$, $d\geq 2$, and $\mu>0$. At zero temperature, the carrier density at small doping is definitely determined by the carriers below Fermi energy, which read
		\begin{align}
			n(t,d,\mu,\Lambda)&=g_s\sum_{\kappa,\lambda=\pm}n_{\kappa}^{\lambda}(t,d,\mu,\Lambda)
			=\frac{g_{s}}{(2\pi)^{d}}\sum_{\kappa,\lambda=\pm}\int_{|\boldsymbol{k}|\le\Lambda}
			\text{d}^{d}\boldsymbol{k}\Theta\left[\mu-\varepsilon_{\kappa}^{\lambda}(\boldsymbol{k})\right]
			\nonumber\\&
			 =\frac{g_{s}g_{v}}{(2\pi)^{d}}\left[\mathcal{K}_{+}(t,d,\mu,\Lambda)+\mathcal{K}_{-}(t,d,\mu,\Lambda)\right],
		\end{align}
		where $\Lambda<+\infty$ denotes an ultraviolet cutoff for open Fermi surface, $g_v=2$ represents the valley degeneracy (we focus on the $\kappa=+$ valley hereafter), and
		\begin{align}
			\mathcal{K}_{\lambda}(t,d,\mu,\Lambda)
			=\int_{|\boldsymbol{k}|\le\Lambda}\text{d}^{d}\boldsymbol{k}
			\Theta\left[\mu-\varepsilon_{+}^{\lambda}(\boldsymbol{k})\right].
		\end{align}
		with $\lambda=\pm$ the band index. For the type-II phase ($t>1$) or type-III phase ($t=1$), the open Fermi surface will dramatically influence the carrier density via the Heaviside function imposed by the geometric structure of the Fermi surface. In 2D, for $t>1$, the conduction band ($\lambda=+$) is related to one branch of hyperbolic Fermi surface satisfying $\mu\ge\varepsilon_{\kappa}^{+}(\boldsymbol{k})$, while the valence band ($\lambda=-$) contributes to the other branch of hyperbolic Fermi surface satisfying $\mu\ge\varepsilon_{\kappa}^{-}(\boldsymbol{k})$. By contrast, for $t=1$, the conduction band ($\lambda=+$) is related to the parabolic Fermi surface $\mu\ge\varepsilon_{\kappa}^{+}(\boldsymbol{k})$, while the valence band ($\lambda=-$) does not contribute to the Fermi surface. It is once again stressed that for 3D tilted Dirac bands the Fermi surfaces are the ellipsoid, paraboloid, and hyperboloid by rotating the ellipse, parabola, and hyperbola (Fermi surfaces in 2D tilted Dirac bands) with respect to the tilting direction, respectively.
		
		\subsection{Integration $\mathcal{K}_{\lambda}(t,d,\mu,\Lambda)$ for type-II and type-III phases}
		
		In 2D, the integration $\mathcal{K}_{+}(t>1,d,\mu,\Lambda)$ for the type-II phase can be divided into two parts as
		\begin{align}
			\mathcal{K}_{+}\left(t>1,d=2,\mu,\Lambda\right)&
			=\Theta(\Lambda_{-}-\Lambda)\mathcal{K}_{+}\left(t>1,d=2,\mu,\Lambda<\Lambda_{-}\right)
			+\Theta(\Lambda-\Lambda_{-})\mathcal{K}_{+}\left(t>1,d=2,\mu,\Lambda>\Lambda_{-}\right).
		\end{align}
		
		\begin{figure*}[htbp]
			\centering
			\includegraphics[width=18cm]{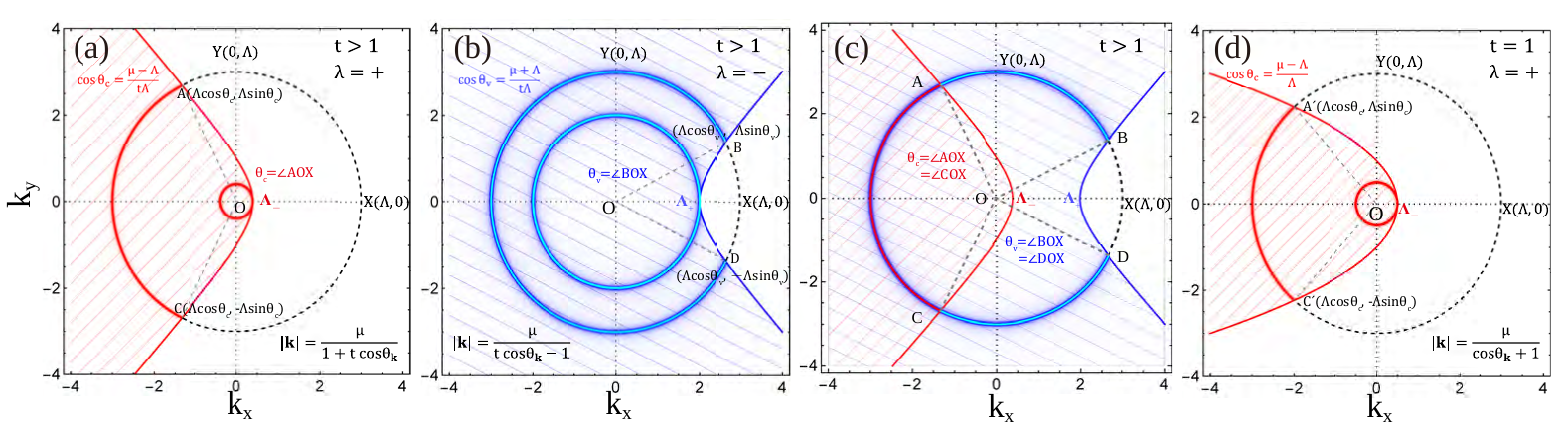}
			\caption{Integration region for calculation of carrier density in 2D type-II and type-III Dirac bands. The Fermi surfaces are hyperbolic in (a)-(c), but parabolic in (d). The integration region is always on the left of hyperbola or parabola. In panel (a), the region is also inside the small halo when the radius $|\boldsymbol{k}|<\Lambda_{-}$ but inside the large halo when the radius $|\boldsymbol{k}|>\Lambda_{-}$. In panel (b), the integration region is also inside the small halo when the radius $|\boldsymbol{k}|<\Lambda_{+}$ but inside the large halo when the radius $|\boldsymbol{k}|>\Lambda_{+}$. In panel (c), the region is also inside the halo when the radius $|\boldsymbol{k}|>\Lambda_{+}$. In panel (d), the integration region is also inside the small halo when the radius $|\boldsymbol{k}|<\Lambda_{-}$ but inside the large halo when the radius $|\boldsymbol{k}|>\Lambda_{-}$.
			}
			\label{figC}
		\end{figure*}
		
		From the integration regions shown in Fig.\ref{figC}(a), one has
		\begin{align}
			\mathcal{K}_{+}\left(t>1,d=2,\mu,\Lambda<\Lambda_{-}\right)&
			 =\int_{|\boldsymbol{k}|\le\Lambda<\Lambda_{-}}\text{d}^{2}\boldsymbol{k}\Theta\left[\mu-\varepsilon_{+}^{+}(\boldsymbol{k})\right]
			\nonumber\\&=
			2\int_{0}^{\pi}\text{d}\theta\int_{0}^{\Lambda}k\text{d}k
			=\pi\Lambda^{2},
		\end{align}
		and
		\begin{align}
			\mathcal{K}_{+}\left(t>1,d=2,\mu,\Lambda>\Lambda_{-}\right)&
			 =\int_{|\boldsymbol{k}|\le\Lambda>\Lambda_{-}}\text{d}^{2}\boldsymbol{k}\Theta\left[\mu-\varepsilon_{+}^{+}(\boldsymbol{k})\right]
			\nonumber\\&= 2\int_{\theta_{c}}^{\pi}\text{d}\theta\int_{0}^{\Lambda}k\text{d}k+2\int_{0}^{\theta_{c}}\text{d}\theta\int_{0}^{{\mu/(t\cos\theta+1)}} k \text{d}k
			\nonumber\\&=
			 \pi\Lambda^{2}-\theta_{c}\Lambda^{2}+\mu^{2}\int_{0}^{\theta_{c}}\frac{\text{d}\theta}{(t\cos\theta+1)^{2}},
		\end{align}
		with $\theta_c=\arccos\frac{\mu/\Lambda-1}{t}$. Finally, we have
		\begin{align}
			\mathcal{K}_{+}\left(t>1,d=2,\mu,\Lambda\right)&
			=\Theta(\Lambda_{-}-\Lambda)\left[\pi\Lambda^{2}\right]
			+\Theta(\Lambda-\Lambda_{-})\left[\pi\Lambda^{2}-\theta_{c}\Lambda^{2}
			+\mu^{2}\int_{0}^{\theta_{c}}\frac{\text{d}\theta}{(t\cos\theta+1)^{2}}\right]
			\nonumber\\&
			=\pi\Lambda^{2}
			-\Theta(\Lambda-\Lambda_{-})\left[\Lambda^{2}\theta_{c}
			-\mu^{2}\int_{0}^{\theta_{c}}\frac{\text{d}\theta}{(t\cos\theta+1)^{2}}\right].
		\end{align}
		
		The integration $\mathcal{K}_{-}(t,d,\mu,\Lambda)$ for the type-II phase in 2D can also be divided into two parts as
		\begin{align}
			\mathcal{K}_{-}\left(t>1,d=2,\mu,\Lambda\right)&
			=\Theta(\Lambda_{+}-\Lambda)\mathcal{K}_{-}\left(t>1,d=2,\mu,\Lambda<\Lambda_{+}\right)
			+\Theta(\Lambda-\Lambda_{+})\mathcal{K}_{-}\left(t>1,d=2,\mu,\Lambda>\Lambda_{+}\right).
		\end{align}
		
		From the integration regions shown in Figs.\ref{figC}(b) and \ref{figC}(c), we have
		\begin{align}
			\mathcal{K}_{-}\left(t>1,d=2,\mu,\Lambda<\Lambda_{+}\right)&
			=\int_{|\boldsymbol{k}|\le\Lambda<\Lambda_{+}}\text{d}^{2}\boldsymbol{k}
			\Theta\left[\mu-\varepsilon_{+}^{-}(\boldsymbol{k})\right]
			\nonumber\\&
			= 2\int_{0}^{\pi}\text{d}\theta\int_{0}^{\Lambda}k\text{d}k
			=\pi\Lambda^{2},
		\end{align}
		and
		\begin{align}
			\mathcal{K}_{-}\left(t>1,d=2,\mu,\Lambda>\Lambda_{+}\right)&
			=\int_{|\boldsymbol{k}|\le\Lambda>\Lambda_{+}}\text{d}^{2}\boldsymbol{k}
			\Theta\left[\mu-\varepsilon_{+}^{-}(\boldsymbol{k})\right]
			\nonumber\\&=2\int_{\theta_{v}}^{\pi}\text{d}\theta\int_{0}^{\Lambda}k\text{d}k
			+2\int_{0}^{\theta_{v}}\text{d}\theta\int_{0}^{\mu/(t\cos\theta-1)}k\text{d}k
			\nonumber\\&=
			 \pi\Lambda^{2}-\theta_{v}\Lambda^{2}+\mu^{2}\int_{0}^{\theta_{v}}\frac{\text{d}\theta}{(t\cos\theta-1)^{2}},
		\end{align}
		with $\theta_v=\arccos\frac{\mu/\Lambda+1}{t}$. Finally, one has
		\begin{align}
			\mathcal{K}_{-}\left(t>1,d=2,\mu,\Lambda\right)&
			=\Theta(\Lambda_{+}-\Lambda)\left[\pi\Lambda^{2}\right]
			+\Theta(\Lambda-\Lambda_{+})\left[\pi\Lambda^{2}-\theta_{v}\Lambda^{2}
			+\mu^{2}\int_{0}^{\theta_{v}}\frac{\text{d}\theta}{(t\cos\theta-1)^{2}}\right]
			\nonumber\\&
			=\pi\Lambda^{2}
			-\Theta(\Lambda-\Lambda_{+})\left[\Lambda^{2}\theta_{v}
			-\mu^{2}\int_{0}^{\theta_{v}}\frac{\text{d}\theta}{(t\cos\theta-1)^{2}}\right].
		\end{align}

		To summarize, the summation $\mathcal{K}_{+}\left(t>1,d=2,\mu,\Lambda\right)+\mathcal{K}_{-}\left(t>1,d=2,\mu,\Lambda\right)$ reads
		\begin{align}
			&\mathcal{K}_{+}\left(t>1,d=2,\mu,\Lambda\right)+\mathcal{K}_{-}\left(t>1,d=2,\mu,\Lambda\right)
			\nonumber\\&
			=2\pi\Lambda^{2}
			-\Theta(\Lambda-\Lambda_{-})\left[\theta_{c}\Lambda^{2}
			-\mu^{2}\int_{0}^{\theta_{c}}\frac{\text{d}\theta}{(t\cos\theta+1)^{2}}\right]
			-\Theta(\Lambda-\Lambda_{+})\left[\theta_{v}\Lambda^{2}
			-\mu^{2}\int_{0}^{\theta_{v}}\frac{\text{d}\theta}{(t\cos\theta-1)^{2}}\right].
		\end{align}
		
		The calculation for the 3D case can be similarly performed by noticing the Fermi surfaces for 3D tilted Dirac bands
		are the ellipsoid, paraboloid, and hyperboloid by rotating the ellipse, parabola, and hyperbola (Fermi surfaces in 2D tilted Dirac bands) with respect to the tilting direction, respectively. Accordingly, the integration $\mathcal{K}_{+}(t>1,d,\mu,\Lambda)$ for type-II phase is given as
		\begin{align}
			\mathcal{K}_{+}\left(t>1,d=3,\mu,\Lambda\right)&
			=\Theta(\Lambda_{-}-\Lambda)\mathcal{K}_{+}\left(t>1,d=3,\mu,\Lambda<\Lambda_{-}\right)
			+\Theta(\Lambda-\Lambda_{-})\mathcal{K}_{+}\left(t>1,d=3,\mu,\Lambda>\Lambda_{-}\right),
		\end{align}
		where
		\begin{align}
			\mathcal{K}_{+}\left(t>1,d=3,\mu,\Lambda<\Lambda_{-}\right)&
			=\int_{|\boldsymbol{k}|\le\Lambda<\Lambda_{-}}\text{d}^{3}\boldsymbol{k}
			\Theta\left[\mu-\varepsilon_{+}^{-}(\boldsymbol{k})\right]
			\nonumber\\&
			=2\pi\int_{0}^{\pi}\sin\theta\text{d}\theta\int_{0}^{\Lambda}k^2\text{d}k
			=\frac{4\pi}{3}\Lambda^{3},
		\end{align}
		and
		\begin{align} \mathcal{K}_{+}\left(t>1,d=3,\mu,\Lambda>\Lambda_{-}\right)&
			=\int_{|\boldsymbol{k}|\le\Lambda>\Lambda_{-}}\text{d}^{3}\boldsymbol{k}
			\Theta\left[\mu-\varepsilon_{+}^{+}(\boldsymbol{k})\right]
			\nonumber\\&
			=2\pi\int_{\theta_{c}}^{\pi}\sin\theta\text{d}\theta\int_{0}^{\Lambda}k^2\text{d}k
			+2\pi\int_{0}^{\theta_{c}}\sin\theta\text{d}\theta\int_{0}^{\mu/(t\cos\theta+1)}k^2\text{d}k
			\nonumber\\&=
			\frac{2\pi}{3}\Lambda^{3}  (1+\cos\theta_{c})+\frac{2\pi}{3}\mu^{3}\int_{0}^{\theta_{c}}\frac{\sin\theta\text{d}\theta}{(t\cos\theta+1)^{3}},
		\end{align}
		with $\theta_c=\arccos\frac{\mu/\Lambda-1}{t}$. From these expressions, we have
		\begin{align}
			\mathcal{K}_{+}\left(t>1,d=3,\mu,\Lambda\right)&
			=\Theta(\Lambda_{-}-\Lambda)\left[\frac{4\pi}{3}\Lambda^{3}\right]
			+\Theta(\Lambda-\Lambda_{-})\frac{2\pi}{3}\left[\Lambda^{3} (1+\cos\theta_{c})+\mu^{3}\int_{0}^{\theta_{c}}\frac{\sin\theta\text{d}\theta}{(t\cos\theta+1)^{3}}\right]
			\nonumber\\&
			=\frac{4\pi}{3}\Lambda^{3}
			-\Theta(\Lambda-\Lambda_{-})\frac{2\pi}{3}\left[\Lambda^{3} (1-\cos\theta_{c})-\mu^{3}\int_{0}^{\theta_{c}}\frac{\sin\theta\text{d}\theta}{(t\cos\theta+1)^{3}}\right].
		\end{align}
		
		In 3D, the integration $\mathcal{K}_{-}(t>1,d,\mu,\Lambda)$ for the type-II phase can be similarly performed as
		\begin{align}
			\mathcal{K}_{-}\left(t>1,d=3,\mu,\Lambda\right)&
			=\Theta(\Lambda_{+}-\Lambda)\mathcal{K}_{-}\left(t>1,d=3,\mu,\Lambda<\Lambda_{+}\right)
			+\Theta(\Lambda-\Lambda_{+})\mathcal{K}_{-}\left(t>1,d=3,\mu,\Lambda>\Lambda_{+}\right),
		\end{align}
		where
		\begin{align}
			\mathcal{K}_{-}\left(t>1,d=3,\mu,\Lambda<\Lambda_{+}\right)&
			=\int_{|\boldsymbol{k}|\le\Lambda<\Lambda_{+}}\text{d}^{3}\boldsymbol{k}
			\Theta\left[\mu-\varepsilon_{+}^{-}(\boldsymbol{k})\right]
			\nonumber\\&
			=2\pi\int_{0}^{\pi}\sin\theta\text{d}\theta\int_{0}^{\Lambda}k^2\text{d}k
			=\frac{4\pi}{3}\Lambda^{3},
		\end{align}
		and
		\begin{align}
			\mathcal{K}_{-}\left(t>1,d=3,\mu,\Lambda>\Lambda_{+}\right)&
			 =\int_{|\boldsymbol{k}|\le\Lambda>\Lambda_{+}}\text{d}^{3}\boldsymbol{k}\Theta\left[\mu-\varepsilon_{+}^{-}(\boldsymbol{k})\right]
			\nonumber\\&
			=2\pi\int_{\theta_{v}}^{\pi}\sin\theta\text{d}\theta\int_{0}^{\Lambda}k^2\text{d}k
			+2\pi\int_{0}^{\theta_{v}}\sin\theta\text{d}\theta\int_{0}^{\mu/(t\cos\theta-1)}k^2\text{d}k
			\nonumber\\&
			=\frac{2\pi}{3}\Lambda^{3}(1+\cos\theta_{v})
			+\frac{2\pi}{3}\mu^{3}\int_{0}^{\theta_{v}}\frac{\sin\theta\text{d}\theta}{(t\cos\theta-1)^{3}},
		\end{align}
		with $\theta_v=\arccos\frac{\mu/\Lambda+1}{t}$. It can be directly obtained that
		\begin{align}
			\mathcal{K}_{-}\left(t>1,d=3,\mu,\Lambda\right)&
			=\Theta(\Lambda_{+}-\Lambda)\left[\frac{4\pi}{3}\Lambda^{3}\right]
			+\Theta(\Lambda-\Lambda_{+})\left[\frac{4\pi}{3}\Lambda^{3}-\frac{2\pi}{3}\Lambda^{3}(1-\cos\theta_{v})
			+\frac{2\pi}{3}\mu^{3}\int_{0}^{\theta_{v}}\frac{\sin\theta\text{d}\theta}{(t\cos\theta-1)^{3}}\right]
			\nonumber\\&
			=\frac{4\pi}{3}\Lambda^{3}
			-\Theta(\Lambda-\Lambda_{+})\frac{2\pi}{3}\left[\Lambda^{3}(1-\cos\theta_{v})
			-\mu^{3}\int_{0}^{\theta_{v}}\frac{\sin\theta\text{d}\theta}{(t\cos\theta-1)^{3}}\right].
		\end{align}
		
		To summarize, the summation $\mathcal{K}_{+}\left(t>1,d=3,\mu,\Lambda\right)+\mathcal{K}_{-}\left(t>1,d=3,\mu,\Lambda\right)$ reads
		\begin{align}
			&\mathcal{K}_{+}\left(t>1,d=3,\mu,\Lambda\right)+\mathcal{K}_{-}\left(t>1,d=3,\mu,\Lambda\right)
			\nonumber\\&
			=\frac{8\pi}{3}\Lambda^{3}
			-\Theta(\Lambda-\Lambda_{-})\frac{2\pi}{3}\left[\Lambda^{3} (1-\cos\theta_{c})-\mu^{3}\int_{0}^{\theta_{c}}\frac{\sin\theta\text{d}\theta}{(t\cos\theta+1)^{3}}\right]
			\nonumber\\&
			-\Theta(\Lambda-\Lambda_{+})\frac{2\pi}{3}\left[\Lambda^{3}(1-\cos\theta_{v})
			-\mu^{3}\int_{0}^{\theta_{v}}\frac{\sin\theta\text{d}\theta}{(t\cos\theta-1)^{3}}\right].
		\end{align}

		For the type-III phase, the corresponding integration can be performed in a similar way. It is noted that the valence band
		always lies below the Fermi energy, namely, $\varepsilon_{\kappa}^{-}(\boldsymbol{k})\leq0<\mu$ for any $\boldsymbol{k}$, which leads to
		\begin{align}
			\Lambda_{+}=\lim_{t\to 1^{+}}\frac{\mu}{|1-t|}=+\infty.
		\end{align}
		Hence we have
		\begin{align}
			\mathcal{K}_{-}\left(t=1,d,\mu,\Lambda\right)
			=\begin{cases}
				\pi\Lambda^{2}, & d=2,\\
				\\
				\frac{4\pi}{3}\Lambda^{3}, & d=3.
			\end{cases}
		\end{align}
		
		For the conduction band, the critical cutoff $\Lambda_{-}=\mu/2$ must be introduced to calculate $\mathcal{K}_{+}\left(t=1,d,\mu,\Lambda\right)$ as
		\begin{align}
			\mathcal{K}_{+}\left(t=1,d,\mu,\Lambda\right)
			=\Theta(\Lambda_{-}-\Lambda)\mathcal{K}_{+}\left(t=1,d,\mu,\Lambda<\Lambda_{-}\right)
			+\Theta(\Lambda-\Lambda_{-})\mathcal{K}_{+}\left(t=1,d,\mu,\Lambda>\Lambda_{-}\right).
		\end{align}
		From the corresponding integration regions shown in Fig.\ref{figC}(d), we obtain that
		\begin{align}
			\mathcal{K}_{+}\left(t=1,d,\mu,\Lambda<\Lambda_{-}\right)
			=\begin{cases}
				\pi\Lambda^{2}, & d=2,\\
				\\
				\frac{4\pi}{3}\Lambda^{3}, & d=3,
			\end{cases}
		\end{align}
		and
		\begin{align}
			\mathcal{K}_{+}\left(t=1,d,\mu,\Lambda>\Lambda_{-}\right)
			=\begin{cases}
				(\pi-\theta_{c})\Lambda^{2}+\mu^{2}\int_{0}^{\theta_{c}}\frac{\text{d}\theta}{(\cos\theta+1)^{2}}, & d=2,\\
				\\
				\frac{2\pi}{3}\Lambda^{3}(1+\cos\theta_{c})+\frac{2\pi}{3}\mu^{3}
				\int_{0}^{\theta_{c}}\frac{\sin\theta\text{d}\theta}{(1+\cos\theta)^{3}}, & d=3,
			\end{cases}
		\end{align}
		with $\theta_c=\arccos\left({\mu}/{\Lambda}-1\right)$, and
		\begin{align}
			\mathcal{K}_{+}\left(t=1,d,\mu,\Lambda\right)
			=\begin{cases}
				\pi\Lambda^{2}-\Theta(\Lambda-\Lambda_{-})\left[\theta_{c}\Lambda^{2}
				-\mu^{2}\int_{0}^{\theta_{c}}\frac{\text{d}\theta}{(\cos\theta+1)^{2}}\right], & d=2,\\
				\\
				\frac{4\pi}{3}\Lambda^{3}-\Theta(\Lambda-\Lambda_{-})\left[\frac{2\pi}{3}\Lambda^{3}(1-\cos\theta_{c})
				-\frac{2\pi}{3}\mu^{3}
				\int_{0}^{\theta_{c}}\frac{\sin\theta\text{d}\theta}{(1+\cos\theta)^{3}}\right], & d=3,
			\end{cases}
		\end{align}
		
		Therefore, the final expression of $\mathcal{K}_{+}\left(t=1,d,\mu,\Lambda\right)+\mathcal{K}_{-}\left(t=1,d,\mu,\Lambda\right)$ takes
		\begin{align}
			&\mathcal{K}_{+}\left(t=1,d,\mu,\Lambda\right)+\mathcal{K}_{-}\left(t=1,d,\mu,\Lambda\right)\nonumber\\
			&=\begin{cases}
				2\pi\Lambda^{2}-\Theta(\Lambda-\Lambda_{-})\left[\theta_{c}\Lambda^{2}
				-\mu^{2}\int_{0}^{\theta_{c}}\frac{\text{d}\theta}{(\cos\theta+1)^{2}}\right], & d=2,\\
				\\
				\frac{8\pi}{3}\Lambda^{3}-\Theta(\Lambda-\Lambda_{-})\left[\frac{2\pi}{3}\Lambda^{3}(1-\cos\theta_{c})
				-\frac{2\pi}{3}\mu^{3}
				\int_{0}^{\theta_{c}}\frac{\sin\theta\text{d}\theta}{(1+\cos\theta)^{3}}\right], & d=3.
			\end{cases}
		\end{align}

		\subsection{Final analytical expressions}
		
		After analytically performing the integrations over $\theta$, the carrier density for type-II and type-III phases ($t\ge 1$) can be explicitly written in a compact form as
		\begin{align}
			n(t\ge 1,d,\mu,\Lambda)&
			=n_{0}(d,\Lambda)
			-\Theta\left(\Lambda-\Lambda_{-}\right)\delta n_{+}(t,d,\mu,\Lambda)
			-\Theta\left(\Lambda-\Lambda_{+}\right)\delta n_{-}(t,d,\mu,\Lambda)
			\nonumber\\&
			=n_{0}(d,\Lambda)
			-\Theta\left[t-\frac{\mu-\Lambda}{\Lambda}\right]
			\delta n_{+}(t,d,\mu,\Lambda)
			-\Theta\left[t-\frac{\mu+\Lambda}{\Lambda}\right]
			\delta n_{-}(t,d,\mu,\Lambda),
		\end{align}
		where
		\begin{align}
			n_{0}(d,\Lambda)=\begin{cases}
				\frac{2\Lambda^{2}}{\pi}, & d=2,\\
				\\
				\frac{4\Lambda^{3}}{3\pi^{2}}, & d=3,
			\end{cases}
		\end{align}
		and
		\begin{align}
			\delta n_{\pm}(t,d,\mu,\Lambda)=\begin{cases}
				 \frac{\Lambda^{2}}{\pi^{2}}\arccos\frac{\frac{\mu}{\Lambda}\mp1}{t}+\frac{\mu^{2}}{\pi^{2}}\mathcal{I}_{\text{2D}}(\frac{\frac{\mu}{\Lambda}\mp1}{t},\pm t), & d=2,\\
				\\
				 \frac{\Lambda^{3}}{3\pi^{2}}\left(1-\frac{\frac{\mu}{\Lambda}\mp1}{t}\right)\pm\frac{\mu^{3}}{3\pi^{2}}\mathcal{I}_{\text{3D}}(\frac{\frac{\mu}{\Lambda}\mp1}{t},\pm t), & d=3,
			\end{cases}
		\end{align}
		with
		\begin{align}
			\mathcal{I}_{\text{2D}}(x,t)=
			\begin{cases} \frac{1}{t^{2}-1}\left[\frac{t\sqrt{1-x^{2}}}{1+tx}-\frac{2}{\sqrt{t^{2}-1}}
				\mathrm{artanh}\sqrt{\frac{(t-1)(1-x)}{(t+1)(1+x)}}\right], & t>1,\\
				\\
				\frac{x+2}{3(x+1)^{2}}\sqrt{1-x^{2}}, & t=1,
			\end{cases}
		\end{align}
		and
		\begin{align}
			\mathcal{I}_{\text{3D}}(x,t)=\frac{1}{2t}\left[\frac{1}{(tx+1)^{2}}-\frac{1}{(t+1)^{2}}\right].
		\end{align}
		
		For $\Lambda<\Lambda_{+}$, one has $\mathcal{I}_{\mathrm{2D}}(x,t)\to \mathcal{I}_{\mathrm{2D}}(x,1)$, $\mathcal{I}_{\mathrm{3D}}(x,t)\to \mathcal{I}_{\mathrm{3D}}(x,1)$, $\Lambda_{+}\to\infty$, and $\Theta\left(\Lambda-\Lambda_{+}\right)\to0$ by setting $t\to1^{+}$. Consequently, one finds that
		\begin{align}
			\lim_{t\to 1^{+}}n(t>1,d,\mu,\Lambda)=n_0(d,\Lambda)
			-\lim_{t\to 1^{+}} \delta n_{+}(t,d,\mu,\Lambda)
			\Theta\left(\Lambda-\Lambda_{-}\right) 	
			=n(t=1,d,\mu,\Lambda),
		\end{align}
		which indicates that $n(t=1,d,\mu,\Lambda)$ can be consistently obtained by taking the limit $t\to 1^{+}$ of $n(t>1,d,\mu,\Lambda)$ when the cutoff $\Lambda<\Lambda_{+}=\mu/(t-1)$. Intuitively, due to the fact that $\Lambda_{+}=\mu/(t-1)$ diverges to $+\infty$ in the limit $t\to1^{+}$, the other branch of hyperbolic Fermi
		surface $\mu\ge\varepsilon_{\kappa}^{-}(\boldsymbol{k})$ for the type-II phase is always outside the region of $|\boldsymbol{k}|\le\Lambda$ contributing to carrier density. As a result, the carrier density for the type-II phase can restore that for the type-III phase.
		
		However, for $\Lambda>\Lambda_{+}=\mu/(t-1)$, one can not obtain $n(t=1,d,\mu,\Lambda)$ by taking the limit $t\to 1^{+}$ of $n(t>1,d,\mu,\Lambda)$, namely,
		\begin{align}
			\lim_{t\to 1^{+}}n(t>1,d,\mu,\Lambda)\neq n(t=1,d,\mu,\Lambda),
		\end{align}
		because the other branch of hyperbolic Fermi surface $\mu\ge\varepsilon_{\kappa}^{-}(\boldsymbol{k})$ for the type-II
		phase also contributes to the carrier density. Furthermore, if $\Lambda\gg\Lambda_{+}=\mu/(t-1)$, we can expand $n(t,d,\mu,\Lambda)$ in a series of $\mu/\Lambda$. Up to the leading order of $\mu$, we have
		\begin{align}
			n(t>1,d,\mu,\Lambda)=\begin{cases}
				\frac{\Lambda^{2}}{\pi}+\frac{\Lambda\mu}{\pi^{2}\sqrt{t^{2}-1}}+\mathcal{O}\left(\mu^{2}\right), & d=2,\\
				\\
				\frac{2\Lambda^{3}}{3\pi^{2}}+\frac{\Lambda^{2}\mu}{\pi^{2}t}+\mathcal{O}\left(\mu^{2}\right), & d=3,
			\end{cases}
		\end{align}
		for the type-II phase ($t>1$), and
		\begin{align}
			n(t=1,d,\mu,\Lambda)=\begin{cases}
				\frac{\Lambda^{2}}{\pi}+\frac{2\Lambda}{3\pi^{2}}\sqrt{2\mu\Lambda}+\mathcal{O}\left(\mu^{3/2}\right), & d=2,\\
				\\
				\frac{2\Lambda^{3}}{3\pi^{2}}+\frac{\Lambda^{2}\mu}{3\pi^{2}}+\mathcal{O}\left(\mu^{2}\right), & d=3,
			\end{cases}
		\end{align}
		for the type-III phase ($t=1$).

	\end{widetext}

\end{document}